\documentclass[journal=jacsat,manuscript=article]{achemso}
\usepackage{chemformula} 
\usepackage[T1]{fontenc} 
\usepackage{epsfig}
\usepackage{graphicx}
\usepackage{titlesec}
\usepackage{subcaption}
\usepackage{dcolumn}
\usepackage{amsthm,amsmath}
\usepackage{comment}
\usepackage[colorlinks]{hyperref}
\usepackage{xcolor}
\usepackage[T1]{fontenc}
\usepackage{mathrsfs}
\usepackage{longtable}
\usepackage{orcidlink}
\usepackage{amsmath}
\usepackage{tabularx}
\usepackage{tabularray}
\usepackage{threeparttable}
\usepackage{caption}
\usepackage[utf8]{inputenc}
\usepackage[T1]{fontenc}
\usepackage[normalem]{ulem}
\usepackage{amssymb}
\usepackage{threeparttable}

\usepackage{graphicx}
\usepackage{dcolumn}
\usepackage{bm}

\usepackage[utf8]{inputenc}
\usepackage[T1]{fontenc}
\usepackage{mathptmx}
\usepackage{etoolbox}
\usepackage{xcolor} 
\usepackage[normalem]{ulem} 
\usepackage{subcaption} 
\usepackage{float}
\usepackage{hyperref}
\usepackage{lmodern}

\usepackage{resizegather}
\UseRawInputEncoding

\author{Sudipta Chakraborty}
\affiliation[Unknown University]
{\small Department of Chemistry, Indian Institute of Technology Bombay, Powai, Mumbai 400076, India}
\author{Muskan Begom}
\affiliation[Unknown University]
{\small Department of Chemistry, Indian Institute of Technology Bombay, Powai, Mumbai 400076, India}
\author{Xubo Wang}
\affiliation[Unknown University]
{\small Department of Chemistry, the Johns Hopkins University, Baltimore, Maryland 21218, United States}
\author{Achintya Kumar Dutta}
\email{achintya@chem.iitb.ac.in}
\affiliation[Unknown University]
{\small Department of Chemistry, Indian Institute of Technology Bombay, Powai, Mumbai 400076, India}

\title[An \textsf{achemso} demo]
  {\Large Efficient Implementation of Relativistic Coupled Cluster Linear Response Theory in Combination with Perturbation Sensitive Natural Spinors and Cholesky Decomposition Treatment of Two-electron Integrals}


\begin{document}

\maketitle
\newpage
\begin{abstract}
We present an efficient implementation of the low-cost linear-response coupled-cluster singles and doubles (LR-CCSD) method for computing static and frequency-dependent polarizabilities in systems with significant relativistic and electron-correlation effects. The implementation combines X2C-based Hamiltonians (X2CAMF and X2CMP), perturbation-sensitive natural spinors (FNS++), and Cholesky decomposition (CD)- based treatment of two-electron integrals to reduce both the computational and memory demands of relativistic LR-CCSD calculations. Benchmark calculations reveal that X2CMP exhibits more robust behavior than X2CAMF in the presence of highly augmented basis sets. The proposed FNS++CD-X2CMP-LR-CCSD approach reproduces four-component reference values with excellent accuracy across a diverse set of atomic and molecular systems. Additionally, different strategies for constructing the FNS++ basis were assessed, and the averaged-density approach was found to offer a favorable balance between accuracy and computational cost. Across the benchmark systems considered in this work, approximately 70\% of the virtual spinor space can be removed with the  FNS++ approach. The present implementation enables accurate and scalable relativistic response calculations for large molecular systems, as demonstrated by the computation of the static and dynamic polarizabilities of uranium hexafluoride using a triple-zeta basis comprising more than 1,400 basis functions.
\end{abstract} 

\newpage

\section{Introduction}

The development of \textit{ab initio} techniques for calculating response properties has advanced significantly over the last few decades, improving accuracy and broadening their applicability. \cite{datta1995coupled, Hendrik77, Helgaker1990,sekino1984linear, casida1995time,christiansen1995second, christiansen1998response, christiansen1999frequency, hald2003calculation, kobayashi1994calculation,nielsen1980transition, hammond2008coupled} Approaches based on quasi-energy formalisms\cite{rice1991calculation} or the Ehrenfest theorem\cite{dalgaard1980time, olsen1985linear} can be used to broadly categorize current formulations of response properties. Practical applications of response theory rely on approximate electronic structure methods such as Hartree-Fock (HF), density functional theory (DFT), and several wave function-based approaches, such as multiconfiguration self-consistent field (MCSCF), configuration interaction (CI), and coupled-cluster (CC) methods.  Among the various response properties, polarizability is crucial for interpreting physical and material properties, including chemical reactivity, optical properties, and intermolecular interactions. \cite{van2005accurate, liu2006infinite, peng2007making, knecht} However, when considering recent advances and state-of-the-art precision technologies, it becomes evident that an accurate description of polarizability requires both an accurate treatment of electron correlation and well-optimized basis sets with a sufficient number of diffuse functions\cite{Champagne}.

However, advances in this area have largely been confined to the nonrelativistic framework. Relativistic effects must be accounted for to accurately describe the electronic structure of heavy elements and molecules containing them in \textit{ab initio} calculations. In addition to the heavy elements, relativistic effects may also influence the properties of molecules composed exclusively of lighter elements. \cite{pyykko1988relativistic} \\
The so-called four-component (4c) schemes,\cite{dyall2007introduction,saue1997principles, klopper1997multiple, nalewajski1995proceedings, peric1996theoretical, el2005theoretical} which are derived directly from the Dirac equation, are the most rigorous approaches to incorporate relativistic effects. Nevertheless, the computational cost of these approaches is significantly higher than that of the corresponding non-relativistic methods. Only a limited number of studies have reported the calculation of linear response properties incorporating relativistic effects, and work in the relativistic regime, especially low-cost implementations, remains limited. \cite{hammond2009accurate} Saue \textit{et al.} \cite{saue2003linear} and Salek \textit{et al.} \cite{salek2005linear} implemented the relativistic four-component (4c) Dirac-Hartree-Fock (DHF) and density functional theory (DFT) methods for polarizability calculations. The predictive accuracy of relativistic DFT calculations remains sensitive to the choice of exchange-correlation functional, and there is no systematic way to improve the accuracy.\cite{hait2018accurate, burke2012perspective} Wave-function-based electron correlation techniques, on the other hand, offer a more robust framework for accurate predictions, allowing systematic improvement.\cite{salek2005comparison,gauss1998triple,larsen1999polarizabilities} The reference wave function for the ground state is generally constructed by performing a four-component Dirac Fock (4c-DHF)\cite{swirles1935relativistic} calculation. The four-component methods are capable of providing highly accurate predictions; however, these methods are associated with significantly higher computational cost and resource requirements compared to standard non-relativistic approaches. To alleviate this cost, various two-component theories\cite{ hess1986relativistic, van1996relativistic, dyall1997interfacing, nakajima1999new, barysz2001two, liu2009exact, saue2011relativistic1} have been introduced. Among the two-component theories, the exact two-component atomic mean field (X2CAMF) method\cite{liu2018, Zhang2022,knechtExactTwocomponentHamiltonians2022} has attracted significant attention due to its balance between cost and accuracy.  In contrast to the molecular mean-field formulation, the X2CAMF scheme avoids the computation of molecular relativistic 2e integrals. As a result, X2CAMF achieves a substantial reduction in computational cost while retaining the dominant relativistic two-electron effects that arise from the strongly localized nature of small-component wave functions near atomic nuclei. One can further improve the X2CAMF method by using the so-called model potential(MP) approach, which adds a correction to the X2C-1e Hamiltonian, defined as the difference between a model X2C mean-field Fock matrix and the X2C-1e Fock matrix.\cite{van2005accurate,wang2025relativistic}\\
Coupled cluster (CC) theory \cite{vcivzek1966correlation,vcivzek1969use,vcivzek1991origins,paldus2005beginnings,crawford2007introduction} is considered one of the most accurate and reliable methods for systems where a single-reference determinant predominantly describes the electronic structure among all the post-Hartree-Fock methods available. This is because it enables an accurate representation of the electron correlation, ensuring size extensivity along with systematic improvability arising from its exponential wave-function parametrization. The Coupled cluster singles doubles (CCSD) approximation is used extensively for small to moderate-sized molecular systems, and it exhibits a formal computational scaling of $O(N^6)$ where $N$ represents the size of the correlation space. Coupled cluster methods within the relativistic four- and two-component Hamiltonian-based approach have been implemented for ground- and excited-state energies,\cite{eliav1994relativistic, eliav1994open, visscher1995kramers, visscher1996formulation, lee1998spin, nataraj2010general, visscher2001formulation, koulias2019relativistic, liu2021relativistic} as well as first- and higher-order properties. \cite{chakraborty2024spin, shee2016analytic, liu2021analytic, zheng2022geometry, yuan2023frequency, yuan2024formulation}

For heavy elements, the values of static and dynamic polarizabilities strongly depend on the choice of basis sets, necessitating large and high-quality basis sets with an adequate number of diffuse functions. \cite{chakraborty2024spin} However, increasing the basis set quality in relativistic coupled-cluster calculations significantly raises computational expense, as relativistic CCSD calculations are approximately 32 times as expensive as the corresponding non-relativistic variant. In recent years, natural spinors have gained attention as a promising and efficient alternative for lowering the computational cost of relativistic wavefunction-based calculations. \cite{mandal2026third,chamoli2022reduced,surjuse2022low,yuan2022assessing,majee2024reduced} Conventional MP2-based frozen natural spinors (FNSs) do not show consistent convergence with respect to truncation for excitation energies and response properties, as Gomes and colleagues have shown. \cite{yuan2023frequency} Recently, state-specific natural spinors have been employed to calculate electron affinities and excitation energies at the CC and ADC levels.\cite{mukhopadhyay2025reduced, chakraborty2025low} Perturbation-sensitive natural spinors (FNS++) have been implemented by Chakraborty \textit{et al.} for static and dynamic molecular property calculations within the 4c framework.\cite{chakraborty2025low-cost}

Since relativistic molecular calculations cannot exploit spin symmetry and need to store matrix elements that are complex numbers, the storage requirement can become a bottleneck. Additionally, the atomic to molecular integral transformation is one of the most computationally expensive steps in 4c relativistic calculations and can be as costly as the CCSD iteration in some cases. Density-fitting methods such as Cholesky decomposition (CD)\cite{helmich2019relativistic,banerjee2023relativistic, uhlirova2024cholesky} can be used to reduce the explicit storage of relativistic four-centered two-electron integrals. In this work, we present an efficient implementation of relativistic LR-CCSD employing X2CAMF and X2CMP Hamiltonian in conjunction with perturbation-sensitive natural spinors and Cholesky decomposition. The performance of the approach is assessed against four-component reference calculations for a diverse set of atomic and molecular systems. We further examine the influence of basis set augmentation, Cholesky thresholds, and FNS++ truncation strategies on the accuracy and efficiency of computed polarizability.

\section{Theory}
\subsection{The X2CMP and X2CAMF schemes}

The four--component Dirac-Coulomb (DC) Hamiltonian in the second-quantized notation can be written as
\begin{equation}
H^{\text{DC}} = \sum_{pq} h^{4c}_{pq}\, a_p^\dagger a_q
+ \frac{1}{4}\sum_{pqrs} g^{\text{C,SD}}_{pq,rs}\, a_p^\dagger a_q^\dagger a_s a_r
+ \frac{1}{4}\sum_{pqrs} g^{\text{C,SF}}_{pq,rs}\, a_p^\dagger a_q^\dagger a_s a_r,
\label{eq:hdc-exact}
\end{equation}

\noindent where $g^{\text{C,SD}}$ and $g^{\text{C,SF}}$ denote, respectively, the spin-dependent and spin-free parts of the two-electron Coulomb interaction. Under the mean-field approximation one can express Eq.~\eqref{eq:hdc-exact} as
\begin{equation}
H^{\text{DC}} \approx \sum_{pq}\left[h^{4c}_{pq} + \sum_i n_i\, g^{\text{C,SD}}_{pi,qi}\right] a_p^\dagger a_q
+ \frac{1}{4}\sum_{pqrs} g^{\text{C,SF}}_{pq,rs}\, a_p^\dagger a_q^\dagger a_s a_r,
\label{eq:hdc-approx}
\end{equation}

\noindent where $i$ runs over the occupied orbitals of the atomic reference and $n_i$ is the corresponding occupation number.  Within exact two-component (X2C) theory,\cite{liu2009exact} the four-component Dirac equation
\begin{equation}
h^{4c} C^{4c} = E^{4c} S^{4c} C^{4c},
\label{eq:dirac}
\end{equation}

\noindent with
\begin{align}
    C^{4c} = \begin{pmatrix} C^{L} \\ C^{S} \end{pmatrix},
    \quad
    h^{4c} = \begin{pmatrix} h^{LL} & h^{LS} \\ h^{SL} & h^{SS} \end{pmatrix},
    \quad
    S^{4c} = \begin{pmatrix} S^{LL} & 0 \\ 0 & S^{SS} \end{pmatrix}.
\end{align}
is block-diagonalized into electronic and positronic blocks in a single step\cite{dyall1997interfacing,liu2009exact} rather than through the iterative or multi-step transformations characteristic of earlier two-component approaches.\cite{hess1986relativistic, wolf2002generalized, nakajima2000higher, reiher2004exactI, reiher2004exactII, peng2009arbitrary, barysz2001two, barysz1997expectation, barysz2002infinite} Introducing the matrix $X$ that connects the small- and large-component coefficients,
\begin{equation}
X C^{L} = C^{S},
\end{equation}
the electronic block of the X2C Hamiltonian is given by
\begin{align}
h^{\text{X2C}}_{+} &= R^{\dagger}\, L^{\text{NESC}}\, R \\
L^{\text{NESC}} &= h^{LL} + X^{\dagger} h^{SL} + h^{LS} X + X^{\dagger} h^{SS} X
\label{eq:x2c}
\end{align}
with the renormalization matrix
\begin{equation}
R = \left[(S^{LL})^{-1}\tilde{S}\right]^{-1/2}, \\
\tilde{S} = S^{LL} + X^{\dagger} S^{SS} X
\end{equation}
chosen in the form that renders the resulting two-component eigenvalue problem,
\begin{equation}
h^{\text{X2C}}_{+} C^{2c} = E\, S^{LL} C^{2c},
\end{equation}
invariant to unitary rotations among the basis functions. Combining $h^{\text{X2C}}_{+}$, evaluated from the bare one-electron Dirac Hamiltonian, with the ordinary non-relativistic Coulomb interaction defines the X2C-1e scheme. This scheme treats one-electron relativistic effects essentially exactly, and because the scalar component of the resulting ``two-electron picture-change'' (2e-pc) error\cite{van2005accurate, samzow1992two, ikabata2021picture} is small, its spin-free variant (SFX2C-1e)\cite{dyall2001interfacing,liu2009exact, cheng2011analytic} serves as a standard and reliable treatment of scalar relativity. The error due to the spin-dependent component is, however, not negligible whenever spin--orbit coupling contributes significantly to the property under study, and recovering it without the cost of a full four-component treatment is the problem addressed by the model-potential and atomic-mean-field constructions discussed below.

The natural reference point for how much of the spin-dependent 2e-pc error must be recovered is the X2C molecular mean-field (X2CMMF) construction.\cite{liu2006infinite, sikkema2009molecular} Here the one-electron Dirac Hamiltonian is first augmented with a mean-field two-electron term,
\begin{equation}
h^{4c,\text{MF}} = h^{4c,\text{1e}} + h^{4c,\text{MF2e}},
\label{eq:mfham}
\end{equation}
whose blocks, restricted to the instantaneous Coulomb (C) interaction in a Hartree--Fock treatment, are
\begin{align}
h^{LL,\text{MF2e}} &= J^{LL,C} - K^{LL,C},\\
h^{LS,\text{MF2e}} &= -K^{LS,C}, \\
h^{SL,\text{MF2e}} &= -K^{SL,C},\\
h^{SS,\text{MF2e}} &= J^{SS,C} - K^{SS,C}.
\label{eq:mfblocks}
\end{align}

The resulting four-component Fock matrix is then decoupled into the two-component representation for the full molecule. X2CMMF recovers the four-component Dirac--Coulomb (DC) result essentially exactly. However, the molecular relativistic two-electron integrals must still be evaluated at each self-consistent-field (SCF) iteration, and the method is therefore exactly as expensive as the parent four-component calculation and offers no computational advantage over it at the DHF level.

The route around this expense rests on the same observation already exploited in Eq~\eqref{eq:hdc-approx}. The small-component wave functions responsible for these costly integrals are sharply localized around the nuclei, so that their contribution to the four-component Fock matrix changes little on going from an isolated atom to that same atom embedded in a molecule. This is the physical basis of the model-potential (MP) construction. Instead of solving for $h^{4c,\text{MF2e}}$ molecule by molecule, one evaluates an X2C mean-field Fock matrix and an X2C-1e Fock matrix using atomic (``model'') density matrices and forms their difference,
\begin{equation}
h^{\text{MP}} = h^{\text{X2CMF(MP)}} - h^{\text{X2C-1e(MP)}},
\label{eq:hMP}
\end{equation}
which is added as a correction to the X2C-1e Hamiltonian,\cite{wang2025relativistic}
\begin{equation}
h^{\text{X2CMP}} = h^{\text{X2C-1e}} + h^{\text{MP}}.
\label{eq:hX2CMP}
\end{equation}

Because both terms in eq \eqref{eq:hMP} are constructed from atomic density matrices, no molecular relativistic two-electron integrals over the small component are required; only a modest set of relativistic integrals, extending to at most three centers and evaluated over uncontracted basis functions, is needed to bring the atomic information into the molecular basis. In the atomic limit this construction reproduces X2CMMF exactly, and it approaches the X2CMMF result for molecules as the atomic model density matrices approach the true molecular ones.

A further simplification restricts all relativistic two-electron integrals entering Eq.~\eqref{eq:hMP} to a single center, eliminating multi-center relativistic integrals altogether. This is the atomic-mean-field (AMF) approximation, and its X2C realization, X2CAMF.\cite{liu2018, zhangAtomicMeanFieldApproach2022} The four-component Dirac--Coulomb interaction is separated into spin-free and spin-dependent contributions, and only the spin-dependent Coulomb (SDC) contribution, contracted over the occupied atomic orbitals, is retained in the AMF correction,
\begin{equation}
h^{\text{MP,X2CAMF}} = h^{\text{X2CMF(MP),SDC}}.
\label{eq:hx2camf}
\end{equation}
and the spin-free contribution, the scalar 2e-pc correction is omitted entirely. However, the scalar-relativistic two-electron integrals are not as localized as the spin-orbit two-electron integrals and their effect can be non-negligible in extended basis sets. 

\subsection{Relativistic linear response coupled cluster method}

Relativistic coupled cluster (CC) theory describes the correlated wave function through an exponential parametrization applied to a reference state,
\begin{equation}
\left| {{\Psi }}_{{CC}} \right\rangle ={{e}^{{\hat{T}}}}\left| {{{\Phi }}_{0}} \right\rangle
\end{equation}
where
\begin{equation}
\hat{T}={{\hat{T}}_{1}}+{{\hat{T}}_{2}}+{{\hat{T}}_{3}}+....+{{\hat{T}}_{n}}
\end{equation}
denotes the cluster operator, and $\left|{\Phi }_{0}\right\rangle$ represents the reference determinant. In second quantization notation, the single and double excitation operators, $\hat{T}_1$ and $\hat{T}_2$, are written as
\begin{align}
\hat{T}_1 &=  \sum\limits_{ia} t_i^a \{ \hat a_a^\dagger \hat a_i \} \\
\hat{T}_2 &= \frac{1}{4} \sum\limits_{ijab} t_{ij}^{ab} \{ \hat a_a^\dagger \hat a_b^\dagger \hat a_j \hat a_i \}
\end{align}
and the general form for an \textit{n}-tuple excitation operator is given by
\begin{equation}
{\hat T_n} = {\left( {\frac{1}{{n!}}} \right)^2}\sum\limits_{ij...ab...}^n {t_{ij...}^{ab...} \{ \hat a_a^\dagger \hat a_b^\dagger...\;{{\hat a}_j}{{\hat a}_i}...\}} 
\end{equation}
Here, $t_{ij...}^{ab...}$ are the cluster amplitudes, while $\hat{a}^\dagger$ and $\hat{a}$ correspond to creation and annihilation operators, respectively. The indices $(i,j,k,...)$ and $(a,b,c,...)$ label occupied and virtual spinors.

The relativistic CCSD energy and cluster amplitudes are obtained by projecting onto the reference and excited determinants, respectively.
\begin{align}
    {{E}_{CC}}=\left\langle  {{\Phi }_{0}} \right|\bar{H}\left| {{\Phi }_{0}} \right\rangle \\
    \left\langle  {{\mu }_{i}} \right|\bar{H}\left| {{\Phi }_{0}} \right\rangle = 0, \quad i=1,2
\end{align}
and 
\begin{align}
    \bar{{H}} = e^{-\hat{T}} \hat{H} e^{\hat{T}}
\end{align}

Here, $\mu_i$ refers to the singly and doubly excited determinants and $\bar{H}$ is the similarity transformed X2CAMF or X2CMP Hamiltonian. Response theory can be formulated based on the coupled cluster framework for property calculation. It focuses on evaluating molecular properties that arise from the interaction of the ground state wave function with an external perturbation. Following time-dependent perturbation theory, the effect of the external field can be added to the unperturbed Hamiltonian as
\begin{align}
    \hat{H} &= \hat{H}_0 +\hat{V}(t) \\
  \hat{V}(t) &= \int_{-\infty}^{\infty} V^{(\omega)} e^{(\alpha-i\omega)t}\, d\omega
\end{align}
where $\hat{H}_0$ is the unperturbed Hamiltonian and $\hat{V}(t)$ denotes the external perturbation or the interaction operator, which vanishes at $t=-\infty$. The $V(\omega)$ is the Fourier transform of $V(t)$ and $\alpha$ represents a real positive infinitesimal quantity, such that $V(-\infty) = 0$. The linear response function for exact states can be expressed as,
\begin{eqnarray}
{\left\langle \left\langle \hat{A}; V^{\omega_1}\right\rangle \right\rangle} =&&\sum\limits_{k}\left[\frac{\left\langle  {{\Psi }_{0}} \right|\hat{A}\left| {{\Psi }_{k}} \right\rangle\left\langle  {{\Psi }_{k}} \right|V^{\omega_1}\left| {{\Psi }_{0}} \right\rangle}{\omega_1 - \omega_k}\right] \nonumber \\
&&-\sum\limits_{k} \left[ \frac{\left\langle  {{\Psi }_{0}} \right|V^{\omega_1}\left| {{\Psi }_{k}} \right\rangle\left\langle  {{\Psi }_{k}} \right|\hat{A}\left| {{\Psi }_{0}} \right\rangle}{\omega_1 + \omega_k} \right]
\label{eq:sos}
\end{eqnarray}
where $\omega_k$ is the excitation energy corresponding to the transition from the ground state $({\Psi }_{0})$ to the $k$-th excited state $({\Psi }_{k})$. The summation runs over all the excited states, and Eq.~(\ref{eq:sos}) is also known as the sum-over-states equation. The calculation of linear response properties using this equation is not practically feasible for larger systems, since it requires the computation of all the excited states.

In CC response theory, the perturbed amplitudes $X_{\mu}^{(1)}$ and $Y_{\mu}^{(1)}$ are the Fourier transforms of the time-dependent CC amplitudes $t_{\mu}^{(1)}$ and $\lambda_{\mu}^{(1)}$, respectively. The expression for solving $X_{\mu}^{(1)}$ is 
\begin{align}
    X_{\mu}^{(1)}(\omega_1+i\alpha) &= \sum_{\nu} \left[-\mathbf{A}+(\omega_1 + i\alpha)\mathbf{I}\right]^{-1}_{\mu\nu}\xi_{\nu}^{(1)}(\omega_1)
\end{align} 
where $\mathbf{I}$ is an identity matrix and $\mathbf{A}$ denotes the Coupled Cluster Jacobian, expressed as
\begin{equation}
    A_{\mu\nu} = \left\langle  {\mu} \right| \left[ \bar{H}_0,\tau_\nu\right] \left| {{\Phi }_{0}} \right\rangle
\end{equation}
and
\begin{equation}
    \xi_{\nu}^{(1)}(\omega_1) = \left\langle  {\nu} \right|\overline{V}{^{(\omega_1)}}\left| {{\Phi }_{0}} \right\rangle
\end{equation}
Similarly for $Y_{\mu}^{(1)}$,
\begin{align}
  && Y_{\mu }^{(1)}({{\omega }_{1}}+i\alpha )\text{ }=-\sum\limits_{\nu }\left({\eta _{\nu }^{(1)}}({{\omega }_{1}})\text{ } +\sum\limits_{\gamma }{{{F}_{\nu \gamma }}}X_{\gamma }^{(1)}({{\omega }_{1}}+i\alpha )\right)  \times \left\{ \mathbf{A}+({{\omega }_{1}}+i\alpha )\mathbf{I} \right\}_{\nu \mu }^{-1}  
\end{align}
where $\eta^{(1)}$ and the matrix $\mathbf{F}$ are defined as
\begin{align}
    &\eta_{\nu}^{(1)}(\omega_1) = \left\langle  {(1 + \hat{\Lambda})} \right|[\,\overline{V}{^{(\omega_1)},\tau_\nu}]\left| {{\Phi }_{0}} \right\rangle, \\
    &F_{\nu\gamma} = \left\langle  {(1 + \hat{\Lambda})} \right|[[\bar{H}_{0}, \tau_\nu],\tau_\gamma ]]\left| {{\Phi }_{0}} \right\rangle .
\end{align}
The $\hat{\Lambda}$ denotes the de-excitation operator, and the linear response function within the CC framework can be defined as,
\begin{align}
    \left\langle \left\langle \hat{\textbf{A}};\hat{\textbf{B}} \right\rangle \right\rangle &= \frac{1}{2} \hat{P}(A,B) \left[ \left\langle \Phi_0 \left| [Y_{\omega_1}^{B}, \bar{A}] \right| \Phi_0 \right\rangle \right. 
     + \left. \left\langle \Phi_0 \left|(1+\hat{\Lambda}) [\bar{A},X_{\omega_1}^{B}] \right| \Phi_0 \right\rangle \right]
\end{align}
Here, $B=V^{\omega_1}$ and the operator $\hat{P}(A, B)$ simultaneously swaps the positions of operators $\hat{A}$ and $\hat{B}$ and applies complex conjugation to the resulting expression. $X_{\omega_1}^{B}$ and $Y_{\omega_1}^{B}$ are the perturbed right and left-hand coupled-cluster amplitudes, respectively, for the operator $\hat{B}$.

\subsection{Natural spinor}

Natural orbitals are defined as the eigenfunctions of the correlated one-body reduced density matrix.\cite{lowdin1955quantum} In a relativistic framework, the analogous quantities, known as natural spinors, are obtained by diagonalizing a spin-coupled one-body reduced density matrix derived from a correlated wave function. 

Within the frozen natural spinor (FNS) approach\cite{chamoli2022reduced,yuan2022assessing}, the occupied spinors are retained at the Hartree-Fock level, while only the virtual space is transformed. The unrelaxed one-body reduced density matrix (RDM) at the MP2 level is given by
\begin{equation}
    \Gamma_{pq} = \left\langle  {\Psi^{(1)}} \right| \{ a_{p}^{\dagger}a_q\} \left| {\Psi^{(1)}} \right\rangle
\end{equation}
where $\left| {{\Psi }^{(1)}} \right\rangle$ is the first-order correction to the Hatree-Fock wave function,
\begin{equation}
    \left| {{\Psi }^{(1)}} \right\rangle = \frac{1}{4}\sum_{ijab} t_{ij}^{ab}\left| \Phi_{ij}^{ab} \right\rangle
\end{equation}
with amplitudes defined as
\begin{equation}
    t_{ij}^{ab} = \frac{\left\langle  {ij} \right|  \left| {ab} \right\rangle}{\epsilon_i + \epsilon_j - \epsilon_a - \epsilon_b}
\end{equation}
Here, $\left\langle  {ij} \right|  \left| {ab} \right\rangle$ denotes anti-symmetrized two-electron integrals, $\epsilon_i$, $\epsilon_a$ are spinor energies, and $\left| \Phi_{ij}^{ab} \right\rangle$ represents doubly excited determinants. The virtual-virtual block of the RDM is expressed as
\begin{equation}
    \Gamma_{ab} = \frac{1}{2}\sum_{ijc}t_{ij}^{ac}t_{ij}^{bc}
\end{equation}

Diagonalization of the RDM,
\begin{equation}
    \boldsymbol{\Gamma} \mathbf{V} = \mathbf{\textit{n}}\mathbf{V}
    \label{eq:rdm_diagonalization}
\end{equation}
yields the natural spinors ($\mathbf{V}$) and their corresponding occupation numbers ($n$). The natural spinor representation typically exhibits increased sparsity, as orbitals with small occupation numbers contribute negligibly to correlation effects. These low-occupation spinors can therefore be truncated without significant loss of accuracy.

Using the truncated set of virtual natural spinors, $\boldsymbol{\tilde{V}}$, the virtual–virtual block of the Fock matrix is transformed as
\begin{equation}
    \mathbf{\tilde{F}}\mathbf{_{vv}}  = \boldsymbol{\tilde{V}}\boldsymbol{^\dagger} \mathbf{F_{vv}} \boldsymbol{\tilde{V}}
    \label{eq:fock_transform}
\end{equation}
and subsequently semi-canonicalized via
\begin{equation}
    \mathbf{\tilde{F}}\mathbf{_{vv}} \mathbf{\tilde{Z}} = \mathbf{\tilde{Z}} \boldsymbol{\tilde{\varepsilon }}
    \label{eq:semi_canonical}
\end{equation}
where $\boldsymbol{\tilde{Z}}$ and $\boldsymbol{\tilde{\varepsilon }}$ denote the semi-canonical spinors and corresponding energies. The final transformation from canonical virtual spinors to the natural spinor basis is performed using
\begin{equation}
\mathbf{B}=\mathbf{\tilde{Z}\tilde{V}}
\label{eq:can_to_fns_transform}
\end{equation}

\subsection{Perturbation sensitive natural spinor}

In analogy to the construction of natural spinors from the ground-state MP2 density, a natural spinor basis for the relativistic LRCCSD can also be generated using a perturbation-dependent one-electron density\cite{chakraborty2025low-cost}. This approach, referred to as FNS++, incorporates information from the external perturbation. Since polarizability is inherently a second-order property, the use of second-order perturbed densities provides a more appropriate and physically meaningful basis compared to ground-state MP2 densities, which do not adequately capture response-related correlation effects.

Within the FNS++ framework, the virtual-virtual block of the second-order one-body reduced density matrix corresponding to a perturbation operator $\hat{A}$ is given by
\begin{align}
    {{[{{\Gamma}^{A}_{ab}}]}^{(2)}} &= \frac{1}{2}{{\left[ t_{ij}^{ac}(A) \right]}^{(1)}}{{\left[ t_{ij}^{bc}(A) \right]}^{(1)}} 
    +{{\left[ t_{i}^{a}(A) \right]}^{(1)}}{{\left[ t_{i}^{b}(A) \right]}^{(1)}}
    \label{eq:2nd_order_density}
\end{align}
where the first-order perturbed amplitudes are defined as
\begin{align}
  & {{\left[ t_{ij}^{ac}(A) \right]}^{(1)}}=\frac{\bar{A}_{ij}^{ac}}{{{{\bar{H}}}_{aa}}+{{{\bar{H}}}_{cc}}-{{{\bar{H}}}_{ii}}-{{{\bar{H}}}_{jj}}+\omega}  \\ 
 & {{\left[ t_{i}^{a}(A) \right]}^{(1)}}=\frac{\bar{A}_{i}^{a}}{{{{\bar{H}}}_{aa}}-{{{\bar{H}}}_{ii}}+\omega}
\end{align}
and
\begin{align}
  & \bar{A}_{ij}^{ac}=\hat{P}(a,c)t_{ij}^{ec}A_{e}^{a}-\hat{P}(i,j)t_{mj}^{ac}{{A}_{i}}^{m} \\ 
 & {{{\bar{H}}}_{ii}}={{F}_{ii}}+\frac{1}{2}t_{in}^{ef}\left\langle \left. in \right\| \right.\left. ef \right\rangle  \\ 
 & {{{\bar{H}}}_{aa}}={{F}_{aa}}-\frac{1}{2}t_{mn}^{fa}\left\langle \left. mn \right\| \right.\left. fa \right\rangle
\end{align}

By substituting the density matrix defined above into Eq.~(\ref{eq:2nd_order_density}) and following the diagonalization and transformation steps outlined in Eqs.~(\ref{eq:rdm_diagonalization}) to (\ref{eq:can_to_fns_transform}), the FNS++ basis can be constructed.

The expressions for the first-order amplitudes can be interpreted as approximations to the full response equations in which only the diagonal elements of the Jacobian matrix $\mathbf{A}$ are retained, and they have previously been shown to provide compact virtual spaces for response properties.\cite{crawford2019reduced} For dipole polarizability calculations, the perturbation operator has different components along the Cartesian directions ($x$, $y$, and $z$). In the present implementation, the perturbation-sensitive density is obtained by averaging the contributions from all three directions. Additionally, one can create a direction-specific density without averaging for each direction. However, it will increase the computational cost as the integral transformation and calculation of ground state $t$ cluster and $\lambda$ amplitudes need to be re-performed for each direction.
It is important to note that the perturbation-dependent one-particle density matrix is not guaranteed to be positive definite.  Therefore, the truncation of the natural spinor space is performed based on the absolute values of the occupation numbers.

\section{Computational Details}
We have implemented the X2CAMF- and X2CMP-based 2c-LRCCSD methods, together with their FNS++ variants, in our in-house software package, BAGH.\cite{dutta2023bagh} BAGH is primarily written in Python, and the computationally intensive parts have been written in Cython and Fortran. BAGH is currently interfaced with PySCF,\cite{pyscf2020, Qiming2015, Qiming2018} GAMESS-US,\cite{Barca2020} socutils\cite{socutils} and DIRAC.\cite{DIRAC_saue2020}
The BAGH software package with the PySCF interface was used to perform all the four-component calculations presented in this work. The X2CAMF and X2CMP calculations were performed with the socutils\cite{socutils} interface to BAGH. The Dyall.vxz(x=2,3,4) basis set has been used for the calculations, and the corresponding augmented versions have been generated using the DIRAC package.\cite{DIRAC_saue2020}  The two-electron integrals in the X2CAMF- and X2CMP-based LR-CCSD calculations are treated using the Cholesky decomposition (CD) technique.\cite{aquilante2011cholesky,folkestad2019efficient,zhang2021toward}  In this work, we employ the conventional single-step algorithm, where Cholesky vectors are formed by an iterative procedure that continues until the largest diagonal element of the ERI matrix falls below the predefined Cholesky threshold ($\tau$). The Cholesky vectors are initially constructed in the atomic orbital (AO) basis and subsequently transformed to the molecular orbital (MO) representation, where antisymmetrized two-electron integrals are formed. Higher-rank integrals involving more than two particle indices are not explicitly built or stored; instead, they are evaluated on the fly using the Cholesky vectors.
Unless otherwise specified, all calculations employed the frozen-core approximation and a Cholesky decomposition threshold of $10^{-5}$.

\section{Results and Discussion}
\subsection{Polarizability of Group IIB atoms}
\begin{figure}[h]
\centering
    \begin{subfigure}{0.8\textwidth} 
        \includegraphics[width=\linewidth]{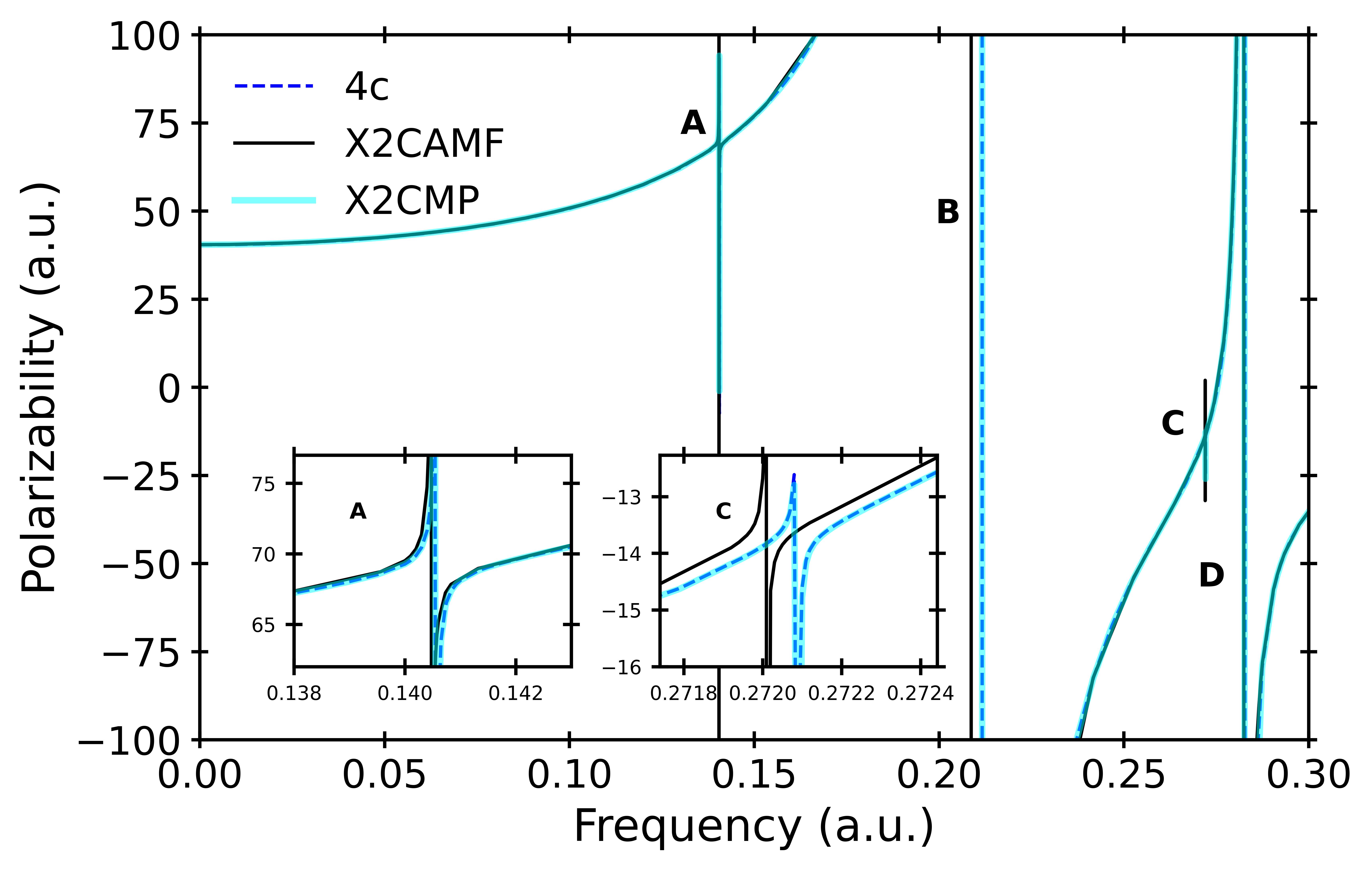} 
        %
    \end{subfigure}

    \caption{\label{fig:zncdhg_dispersion}Polarizability spectra of Zn, using relativistic LRCC using 4c-DC, X2CAMF and X2CMP Hamiltonian with the s-aug-dyall.v2z basis set.}
\end{figure}

To test the accuracy of the X2CAMF and X2CMP-based LR-CCSD method, we have calculated the dynamic polarizability of Zn, Cd, and Hg atoms. These systems have been widely used as benchmarks in previous LRCCSD implementations based on the SFX2C1e,\cite{chakraborty2025spin} X2C,\cite{yuan2024formulation} and four-component (4c)\cite{chakraborty2025low-cost} Hamiltonians, all employing the s-aug-dyall.v2z basis set. From Fig.~\ref{fig:zncdhg_dispersion}, it can be seen that both the X2CMP- and X2CAMF-based LR-CCSD methods exhibit excellent agreement with the 4c data over the entire frequency range for the Zn atom. Analogous behavior is also observed for Cd and Hg, as presented in Figs.~S1 and S2 of the SI. Importantly, the pole positions corresponding to the spin-forbidden ${}^{1}S_{0}\rightarrow{}^{3}P_{1}$ and spin-allowed ${}^{1}S_{0}\rightarrow{}^{1}P_{1}$ transitions are accurately reproduced by both Hamiltonians. Furthermore, X2CMP and X2CAMF Hamiltonians accurately reproduce the widths of the poles for Zn, Cd, and Hg, closely following the trends observed in the 4c calculations. In particular, the systematic increase in pole widths from Zn to Hg, arising from the enhancement of spin-orbit coupling, is clearly reflected in both approaches. These results collectively demonstrate the accuracy of the present X2CMP and X2CAMF-based LRCCSD implementations.

However, closer inspection of resonances A and C, shown in the insets, reveals small but systematic deviations between X2CAMF and the 4c-DC or X2CMP Hamiltonian that are otherwise obscured on the scale of the main panel. Near resonance A (0.138 - 0.143 a.u.), the X2CAMF pole is shifted slightly relative to 4c-DC, while X2CMP remains in close agreement with the four-component value. This discrepancy is markedly more pronounced for resonance C (0.2718 - 0.2724 a.u.), where X2CAMF deviates visibly from 4c-DC, whereas X2CMP continues to track the 4c values closely.  Given that X2CMP and X2CAMF differ only in the treatment of multi-center relativistic two-electron contributions, with X2CAMF retaining exclusively one-center, spin-dependent 2e-PC terms and omitting the scalar 2e-PC contribution entirely, on the grounds that the latter is expected to be short-ranged, the deviation observed for the dynamic polarizability has non-negligible sensitivity in augmented basis sets to the multi-center character of the scalar 2e-PC correction, which is retained in X2CMP but discarded in X2CAMF. These results indicate that both X2CMP and X2CAMF reproduce 4c-DC LRCC polarizabilities for Zn to a high degree of accuracy at substantially reduced integral cost, but that the neglect of the multi-center scalar 2e-PC term in X2CAMF can introduce slight errors. This is consistent with the theoretical basis for the X2CAMF approximation, which is justified specifically for the short-ranged spin-dependent 2e-PC terms and not for the longer-range scalar contribution.

%
%
%
%
%

\subsection{Comparison of FNS and FNS++ }

\begin{figure}
\centering
    \begin{subfigure}{0.48\textwidth} 
        \includegraphics[width=\linewidth]{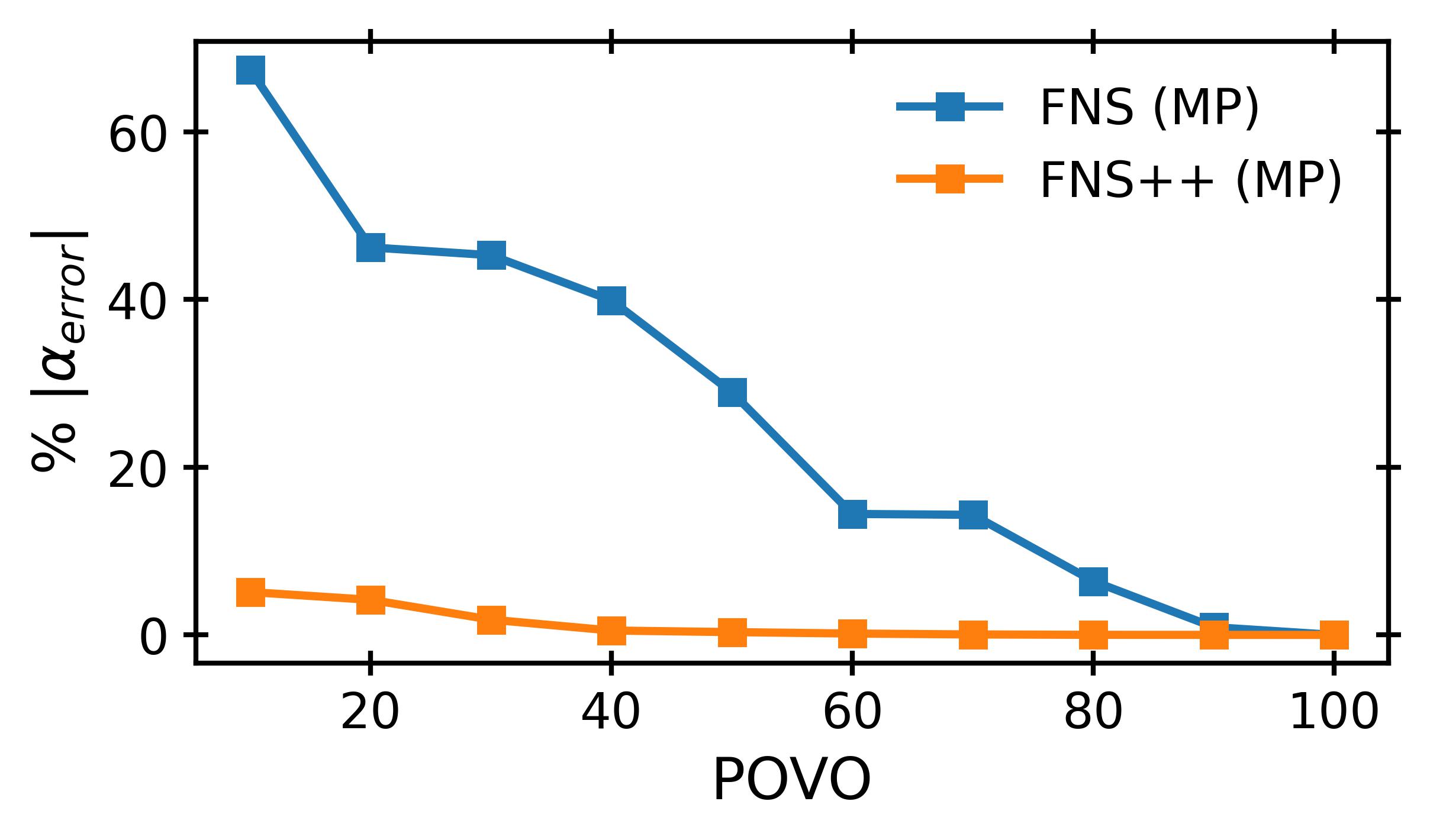} 
        \caption{}
        
    \end{subfigure}
    \begin{subfigure}{0.48\textwidth} 
        \includegraphics[width=\linewidth]{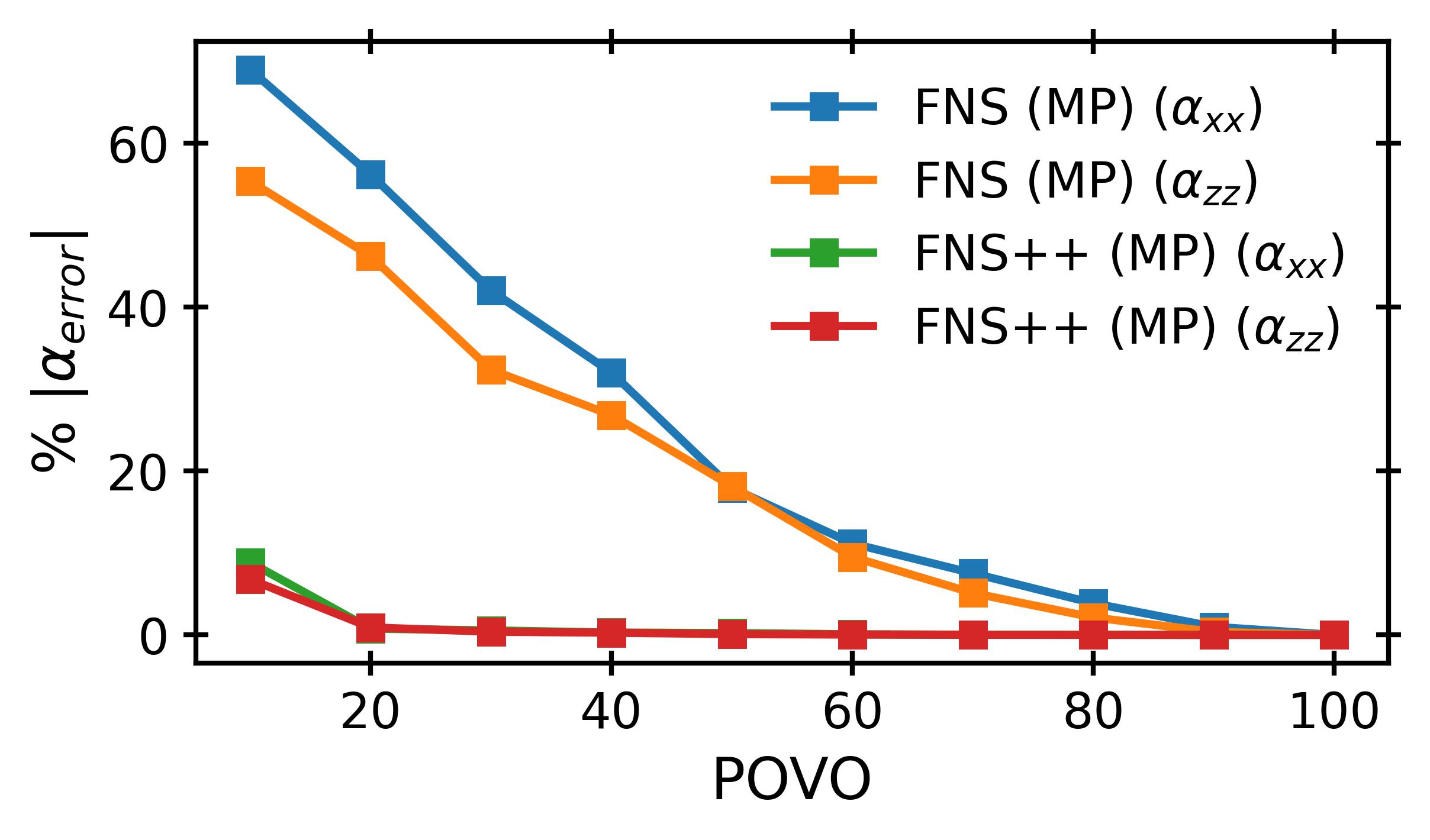} 
        \caption{}
        \label{fig:sub2}
    \end{subfigure}
    
    \caption{\label{fig:povo_zn}Percentage of error in (a) static polarizability of the Zn atom and HBr molecule computed with the uncontracted s-aug-dyall.v2z basis set, using FNS and FNS++ truncation schemes and X2CMP (MP) Hamiltonian, as a function of the percentage of virtual orbitals (POVO) retained.}
\end{figure}

Figure~\ref{fig:povo_zn} illustrates the percentage error in the static polarizability of (a) the Zn atom and (b) the HBr molecule as a function of the percentage of virtual orbitals retained (POVO), using the uncontracted s-aug-dyall.v2z basis set. Results are shown for both the FNS and FNS++ truncation schemes in combination with the X2CMP (MP) Hamiltonian. For the standard FNS truncation, a pronounced dependence on the size of the retained virtual space is observed for the static polarizability. At low POVO values, the errors are substantial, exceeding 60--70\% for the Zn atom and remaining above 40\% even at intermediate truncation levels. Although the error decreases monotonically as more virtual orbitals are included, convergence to within 1--2\% is achieved only when nearly the full virtual space ($\gtrsim 90\%$ POVO) is retained. This behavior highlights the limited efficiency of the FNS scheme in describing response properties, which are inherently sensitive to the nature of the construction of virtual spinors from the ground state relativistic correlated density. In contrast, the FNS++ truncation scheme exhibits dramatically improved convergence behavior. Even at very aggressive truncation levels (POVO $\sim 20$--30\%), the error in the polarizability is already reduced to below 5\%, and it rapidly approaches higher accuracy as POVO increases. Beyond approximately 50\% POVO, the errors associated with FNS++ become essentially negligible, remaining close to zero for the static polarizability. This striking improvement demonstrates the effectiveness of the FNS++ scheme in retaining the most relevant virtual contributions for linear response properties. For the HBr molecule, the same behavior is observed for both the perpendicular ($\alpha_{xx}$) and parallel ($\alpha_{zz}$) components of the polarizability tensor. The trend is consistent with that observed in the previous implementation of FNS++-LR-CCSD within the 4c-DC framework.\cite{chakraborty2025low-cost} The trends observed for the dynamic polarizability closely parallel those found in the static case, indicating that the advantages of the FNS++ truncation scheme are not limited to the zero-frequency limit (See Fig. S3 in SI). At finite frequency, where the response is governed by an explicit interplay between excitation energies and transition moments, an accurate and balanced representation of the virtual orbital space becomes even more critical. The ability of FNS++ to maintain low errors across a wide range of POVO values, therefore, demonstrates that it effectively preserves the essential frequency-dependent contributions to the linear response function. This behavior can be traced back to the density constructed at the canonical level, where the dependence on the external field frequency enters explicitly through the frequency-dependent singles and doubles response amplitudes. The consistent performance across static and frequency-dependent regimes thus demonstrates the general applicability of the FNS++ truncation strategy for frequency-dependent response calculations over a broad frequency range.
\begin{figure}
\centering
    %
    \begin{subfigure}{0.48\textwidth} 
        \includegraphics[width=\linewidth]{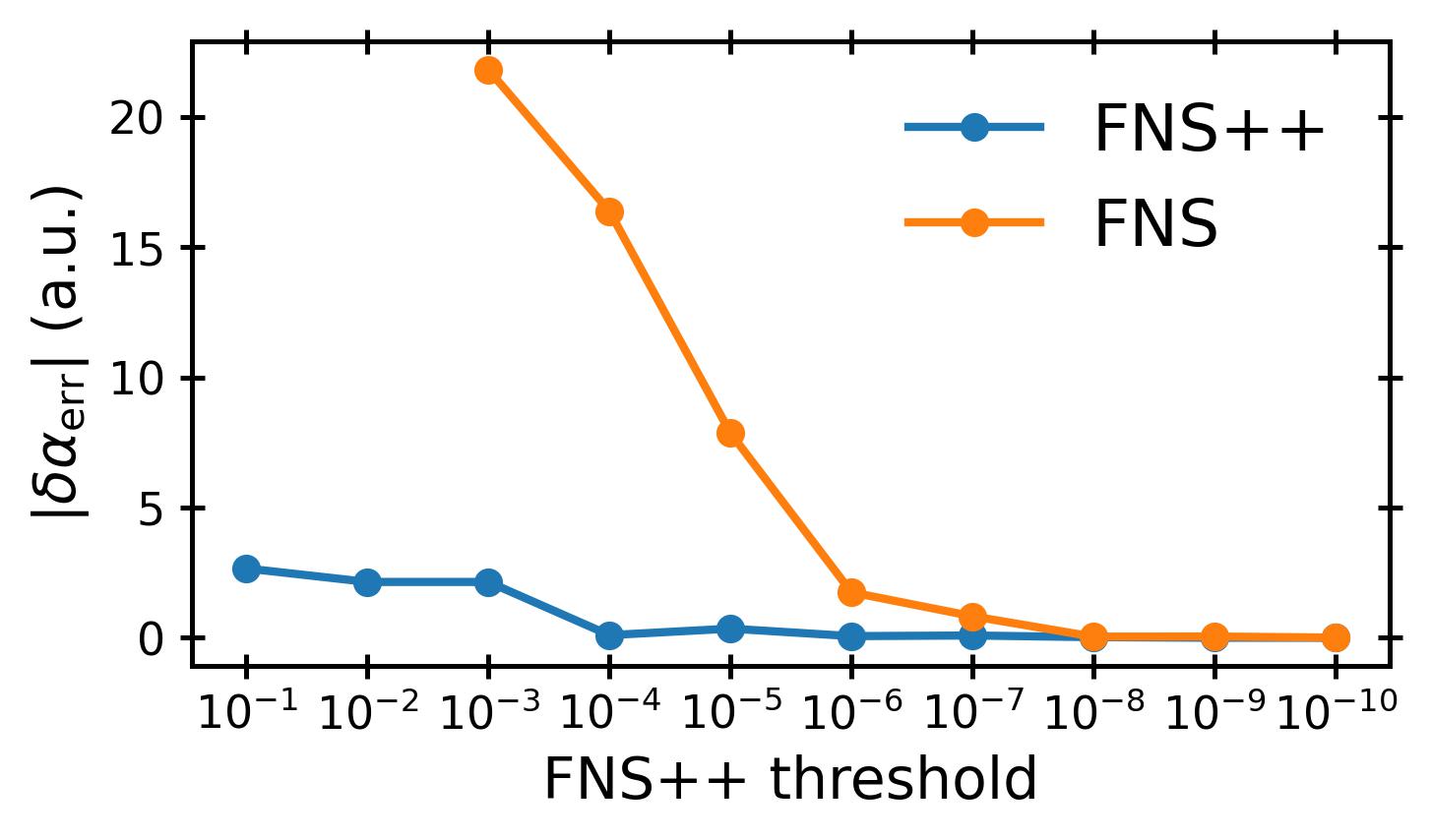} 
        \caption{}
        
    \end{subfigure}
    \begin{subfigure}{0.48\textwidth}
        \includegraphics[width=\linewidth]{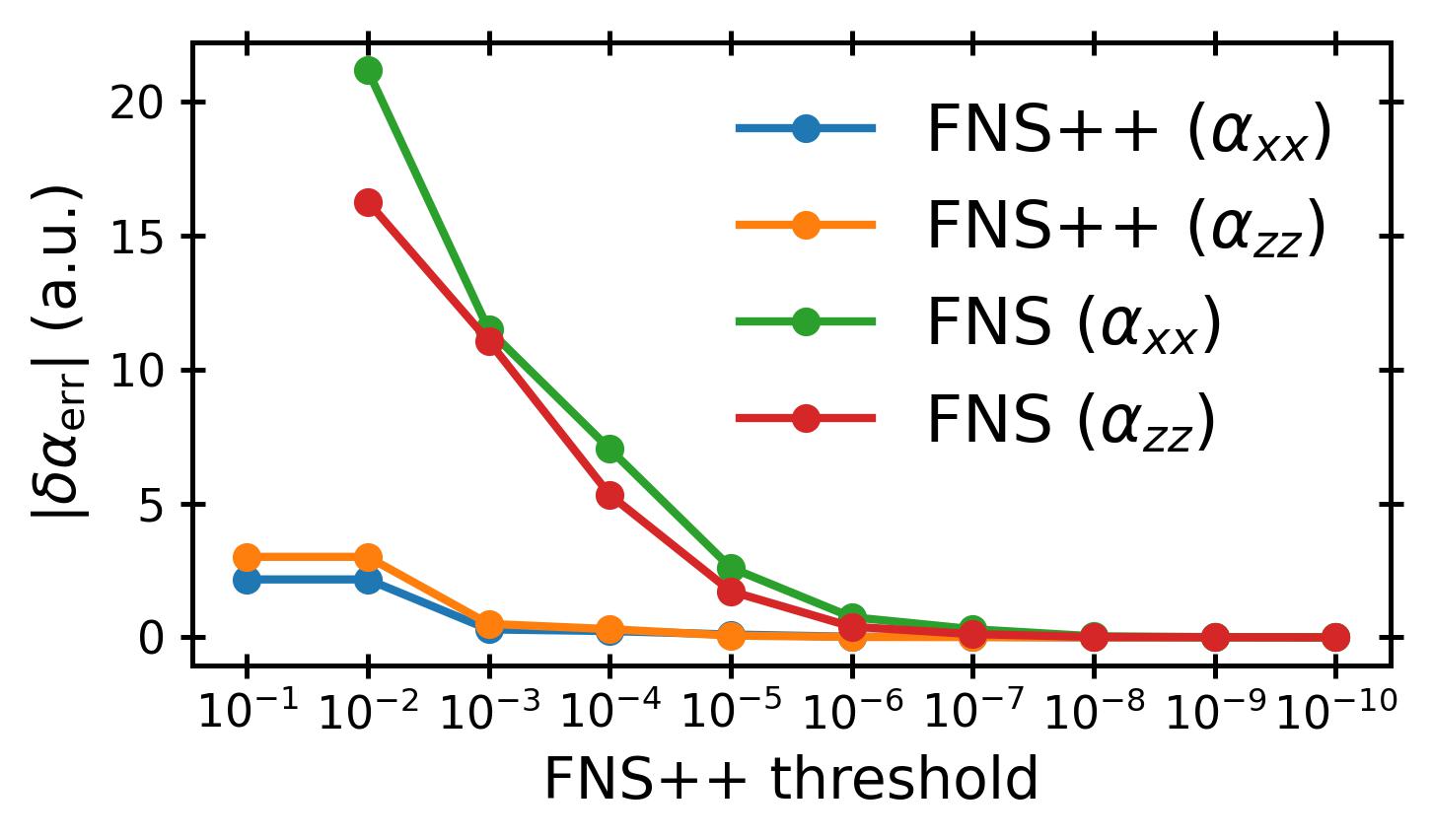}
        \caption{}
        \label{fig:sub2}
    \end{subfigure}
    
    \caption{\label{fig:zn_hbr_fnsthresh}Absolute Error in dynamic ($\omega$ = 0.072 a.u.) polarizability of the Zn atom (a) and static polarizability of the HBr molecule (b) computed with the uncontracted s-aug-dyall.v2z basis set, using FNS and FNS++ truncation schemes and X2CMP Hamiltonians, as a function of the FNS and FNS++ truncation threshold.}
\end{figure}


Since POVO does not provide a reliable criterion for defining an optimal truncation threshold, we next assess the performance of the FNS and FNS++ schemes with the X2CMP Hamiltonian for the polarizability of Zn and HBr as a function of the occupation threshold. Figure~\ref{fig:zn_hbr_fnsthresh} presents the dependence of the dynamic polarizability of Zn at an external frequency of 0.072~a.u. and the static polarizability of HBr on the FNS and FNS++ truncation thresholds. For the Zn atom, owing to its spherical symmetry ($\alpha_{xx}=\alpha_{yy}=\alpha_{zz}$), only the mean polarizability is reported. In contrast, for the anisotropic HBr molecule, two distinct tensor components, $\alpha_{xx}(=\alpha_{yy})$ and $\alpha_{zz}$, are shown. As evident from Fig.~\ref{fig:zn_hbr_fnsthresh}(a), the convergence behavior with respect to the occupation threshold is markedly inferior for the FNS scheme compared to FNS++. In the case of FNS, an absolute error exceeding 20 a.u. is observed at a truncation threshold of $10^{-3}$, whereas the FNS++ scheme yields an error of less than 2.5~a.u. at the same threshold. Upon tightening the threshold to $10^{-4}$, the error in FNS decreases to approximately 15 a.u., while the FNS++ results are already converged to the canonical reference. In fact, the FNS scheme requires an occupation threshold as tight as $10^{-8}$ to achieve satisfactory convergence, highlighting its significantly slower and less efficient convergence behavior. A similar trend is observed for the HBr molecule, where the FNS scheme again exhibits substantial difficulty in converging relative to FNS++. Notably, at relatively loose truncation thresholds, the deviations from the canonical values for both $\alpha_{xx}$ and $\alpha_{zz}$ remain consistent and well-controlled within the FNS++ framework.  On the basis of the above findings, an FNS++ occupation truncation threshold of $10^{-5}$ is chosen as an optimal compromise between accuracy and computational efficiency for all subsequent calculations.

\begin{figure}[h]
\centering
    \begin{subfigure}{0.7\textwidth} 
        \includegraphics[width=\linewidth]{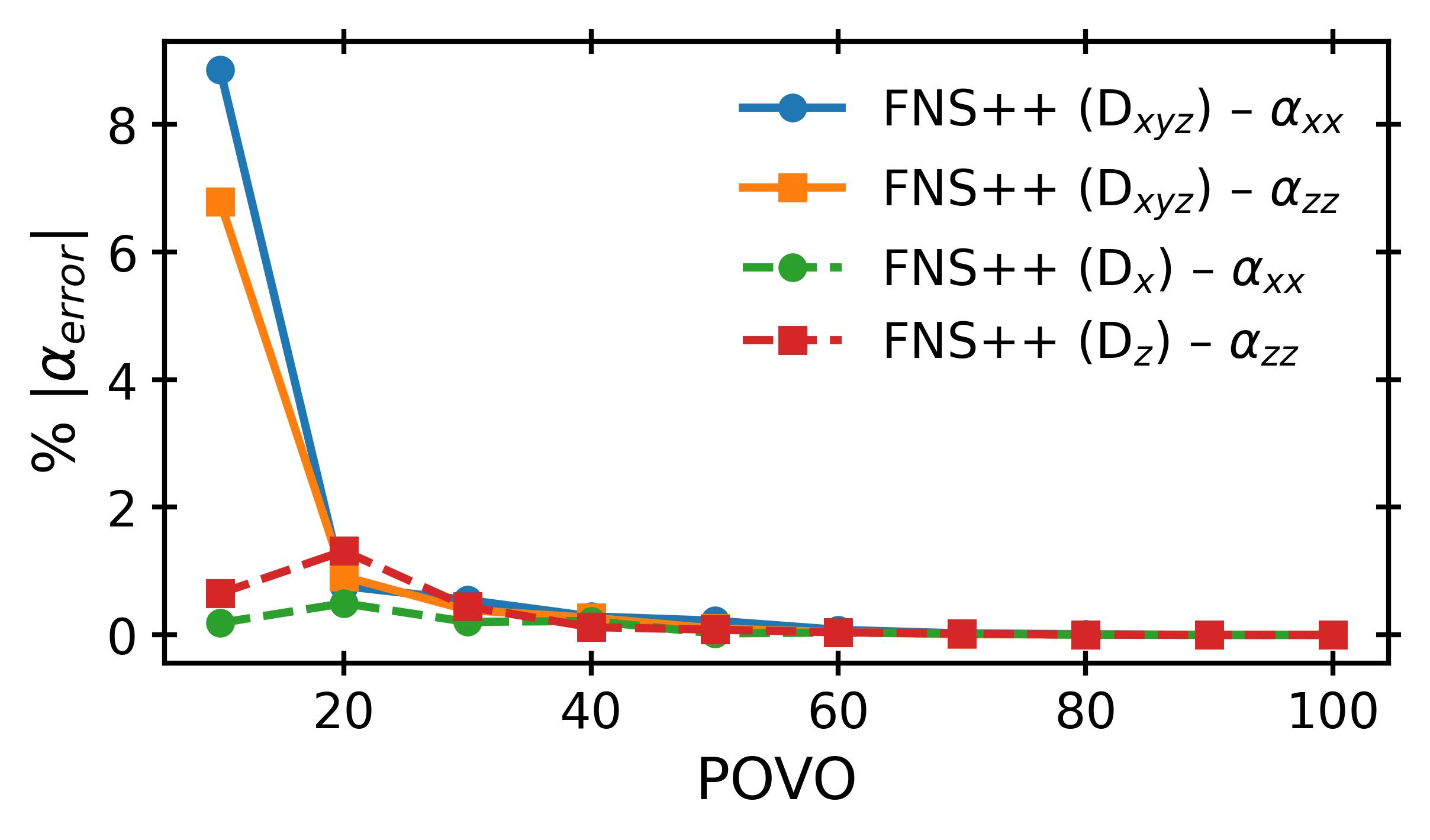} 
        
    \end{subfigure}
    
    \caption{\label{fig:hbr_avg_vs_xz} Static polarizability of the HBr molecule as a function of POVO using average and direction-specific density in the FNS++ basis. s-aug-dyall.v2z basis set is used.}
\end{figure}

The results are also sensitive to how the FNS++ basis is constructed. In the preceding discussion, the density is constructed by considering perturbations along all three Cartesian directions, followed by averaging to obtain a single effective density, from which the FNS++ basis is generated. An alternative approach is to construct the density separately for each Cartesian direction and subsequently evaluate the corresponding component of the polarizability tensor using the direction-specific density. This procedure is repeated independently for each component.
To assess how these two strategies influence the behavior of polarizability components as a function of POVO, we performed a comparative study on the HBr molecule using the s-aug-dyall.v2z basis set, as shown in Figure~\ref{fig:hbr_avg_vs_xz}. The figure presents the variation of the parallel and perpendicular components of the polarizability obtained using the averaged density $D_{xyz}$ and the direction-specific densities $D_x$ and $D_z$.
It is observed that, when a small percentage of the virtual space is retained, the direction-specific approach yields improved accuracy for both $\alpha_{xx}$ and $\alpha_{zz}$ compared to the averaged-density approach. However, beyond a POVO threshold of approximately $20\%$, both methods produce comparable results. Despite this advantage at low POVO, the direction-specific construction is computationally more demanding, as it requires separate integral transformations and ground-state coupled-cluster calculations for each Cartesian direction.
In the present implementation, we therefore adopt the averaged-density approach for most calculations, as it provides a favorable balance between computational efficiency and accuracy.

\subsection{Choice of Cholesky Decomposition Threshold}

\begin{figure}[h]
\centering
    \begin{subfigure}{0.8\textwidth} 
        \includegraphics[width=\linewidth]{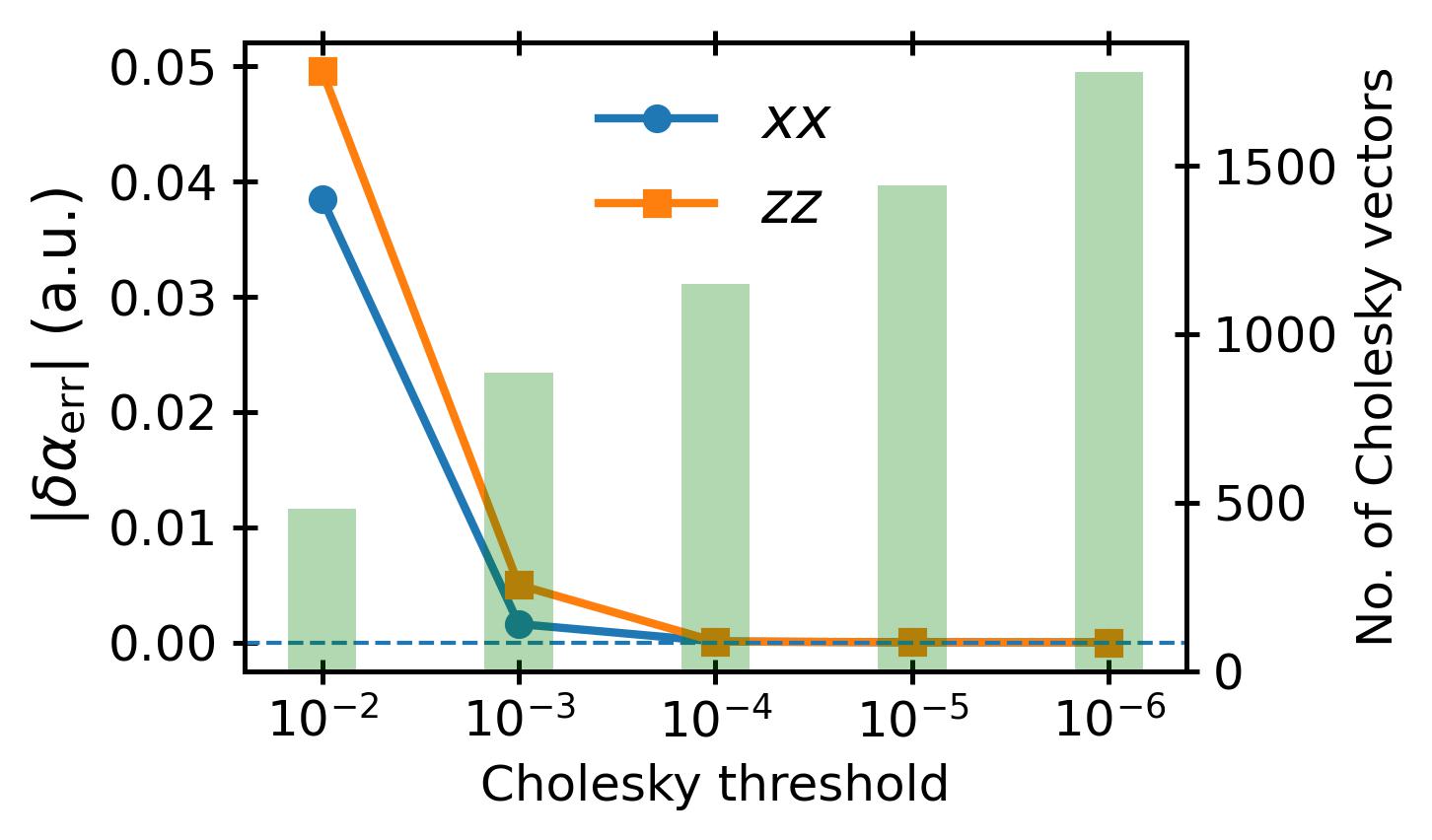} 
        
    \end{subfigure}
    
    \caption{\label{fig:cholesky_thresh_hbr}Absolute error in the static polarizability of the HBr molecule computed with the uncontracted d-aug-dyall.v4z basis set, using a $10^{-5}$ FNS++ truncation threshold and X2CMP Hamiltonians, as a function of the Cholesky threshold. The height of the bars represents the number of Cholesky vectors at each threshold.}
\end{figure}

Since the two-electron integrals in the present relativistic LR-CCSD module are implemented using CD, it is essential to assess the sensitivity of the computed polarizabilities to the choice of the CD threshold.  Figure~\ref{fig:cholesky_thresh_hbr} illustrates the dependence of the absolute error in the static polarizability of the HBr molecule on the Cholesky decomposition threshold, using the uncontracted d-aug-dyall.v4z basis set. An FNS++ truncation threshold of $10^{-5}$ and the X2CMP Hamiltonian have been used. Fig~\ref{fig:cholesky_thresh_hbr} shows the errors in the perpendicular ($\alpha_{xx}=\alpha_{yy}$) and parallel ($\alpha_{zz}$) components of the polarizability tensor. The X2CMP-LR-CCSD results calculated with conventional integrals are taken as the reference. The bars in Fig.~\ref{fig:cholesky_thresh_hbr} indicate the number of Cholesky vectors retained at each threshold. At relatively loose Cholesky thresholds ($10^{-2}$), both the parallel and perpendicular components of the polarizability tensor exhibit noticeable deviations from the reference values, with absolute errors of approximately 0.04--0.05~a.u.
Tightening the threshold to $10^{-3}$ leads to a substantial reduction in the error by nearly an order of magnitude, particularly for the perpendicular component, indicating rapid improvement in the accuracy of the response properties as the quality of the Cholesky representation increases. For thresholds of $10^{-4}$ and tighter, the errors in both $\alpha_{xx}$ and $\alpha_{zz}$ become negligibly small and effectively converge to the result obtained from conventional integrals without CD. Importantly, the convergence behavior of the two tensor components is nearly identical beyond this point, demonstrating that the Cholesky approximation introduces no anisotropic bias in the polarizability tensor when a sufficiently tight threshold is employed. This uniform convergence is essential for reliable prediction of anisotropic response properties in molecular systems. 
The height of the bars further reveals the expected monotonic increase in the number of Cholesky vectors as the threshold is tightened, reflecting the growing computational cost. However, the rapid saturation of the polarizability errors at moderate thresholds indicates that high accuracy can be achieved without resorting to excessively tight Cholesky thresholds. In particular, a threshold of $10^{-4}$ or $10^{-5}$ already provides a satisfactory description of the static polarizability while maintaining a manageable number of Cholesky vectors. Although convergence is achieved at looser thresholds, a Cholesky threshold of $10^{-5}$ is chosen throughout to provide a safe, reliable margin of accuracy.

\begin{table}[ht]
\caption{\label{tab:table_cl2_static}
Basis set benchmarking for the static polarizability of Cl$_2$ using
CD-X2CAMF/X2CMP-FNS++LRCCSD at a $10^{-5}$ FNS++ occupation threshold.
The experimental value is $30.43 \pm 0.30$ a.u.
}

\centering
\begin{tabular*}{\textwidth}{@{\extracolsep{\fill}}l
D{.}{.}{4}D{.}{.}{4}
D{.}{.}{4}D{.}{.}{4}
D{.}{.}{4}D{.}{.}{4}@{}}
\hline
\hline
& \multicolumn{2}{c}{dyall.v2z}
& \multicolumn{2}{c}{dyall.v3z}
& \multicolumn{2}{c}{dyall.v4z} \\
\cline{2-3}\cline{4-5}\cline{6-7}
& \multicolumn{1}{c}{X2CAMF}
& \multicolumn{1}{c}{X2CMP}
& \multicolumn{1}{c}{X2CAMF}
& \multicolumn{1}{c}{X2CMP}
& \multicolumn{1}{c}{X2CAMF}
& \multicolumn{1}{c}{X2CMP} \\
\hline
no-aug & 17.3232 & 17.3234 & 23.9447 & 23.9449 & 27.4588 & 27.4589 \\
s-aug  & 29.5614 & 29.5615 & 30.7665 & 30.7668 & 30.9137 & 30.9117 \\
d-aug  & 30.7082 & 30.7084 & 30.8919 & 30.9008 & 31.0291 & 31.0238 \\
t-aug  & 30.7281 & 30.7285 & 30.9031 & 30.9032 & \multicolumn{1}{c}{---} & 31.0239 \\
q-aug  & 30.7317 & 30.7306 & 30.9145 & 30.9034 & 90.5705 & 31.0235 \\
\hline
\hline
\end{tabular*}
\end{table}

\subsection{Basis set benchmarking}

Table~\ref{tab:table_cl2_static} summarizes the basis-set dependence of the static polarizability of Cl$_2$ obtained using the CD-X2CAMF and X2CMP formulations within the FNS++-LRCCSD framework, keeping FNS++ and Cholesky threshold at 10$^{-5}$ respectively. A pronounced dependence on the inclusion of diffuse functions is observed for all three basis-set qualities. Without augmentation, the computed polarizabilities are severely underestimated, even at the dyall.v4z level, reflecting the essential role of diffuse basis functions in accurately describing the response properties of molecular systems. Upon inclusion of a single set of diffuse functions (s-aug), the polarizability increases substantially and approaches the experimental reference value of $30.43 \pm 0.30$ a.u. Convergence improves with additional augmentation levels, and the static polarizability is effectively converged at the d-aug level for both dyall.v3z and dyall.v4z basis sets. The dyall.v3z and dyall.v4z results with d-aug and higher augmentation are in excellent agreement with experiment, differing by less than 2-3\%.

A close comparison between the X2CAMF and X2CMP Hamiltonians reveals nearly identical polarizabilities across all basis sets without augmentation. However, anomalous behavior is observed with augmentation. The coupled-cluster response equations do not converge for the t-aug dyall.v4z basis set when the X2CAMF Hamiltonian is used. For the q-aug dyall.v4z basis set, although the response equations converge with the X2CAMF Hamiltonian, they lead to an unphysically large polarizability value 90.5705 a.u. The scalar two-electron integrals are not as localized as the corresponding spin-orbit integrals, and the neglect of the two-electron picture-change correction can lead to numerical instability, especially when an extended basis set with a large number of diffuse functions is used. Importantly, such behavior is absent in the corresponding X2CMP results, where the scalar two-electron integrals are treated more accurately using an effective one-electron approximation. Both approximations differ only in the Hartree-Fock step and lead to identical cost for the LR-CCSD step. Therefore, all further calculations in the manuscript are performed using the X2CMP approximation.

\subsection{Benchmarking with four-component results in FNS++ basis}

\begin{table}[ht]
\begin{threeparttable}
\caption{\label{tab:zncdhg_4z}
Static and dynamic polarizabilities (a.u.) of Zn, Cd, and Hg using a
$10^{-5}$ FNS++ occupation threshold with various Hamiltonians and basis sets.
}

\centering
\begin{tabular*}{\textwidth}{@{\extracolsep{\fill}}l
D{.}{.}{5}
D{.}{.}{2}
D{.}{.}{2}
D{.}{.}{2}
D{.}{.}{2}
c@{}}
\hline
\hline
&
\multicolumn{1}{c}{$\omega$} &
\multicolumn{1}{c}{4c\tnote{a}} &
\multicolumn{1}{c}{X2C\tnote{b}} &
\multicolumn{1}{c}{X2CMP\tnote{c}} &
\multicolumn{1}{c}{X2CMP\tnote{d}} &
Expt. \\
\hline

Zn
& 0.00000 & 39.70 & 40.42 & 40.42 & 39.52 &
38.80 $\pm$ 0.80\cite{goebel1996theoretical} \\
& 0.07198 & 44.16 & \multicolumn{1}{c}{---} &
\multicolumn{1}{c}{---} & 43.81 &
43.03 $\pm$ 0.32\cite{goebel1996theoretical} \\
& 0.08383 & 46.05 & \multicolumn{1}{c}{---} &
\multicolumn{1}{c}{---} & 45.67 &
44.76 $\pm$ 0.31\cite{goebel1996theoretical} \\
& 0.14014 & 66.36 & \multicolumn{1}{c}{---} &
\multicolumn{1}{c}{---} & 65.59 &
63.26 $\pm$ 0.12\cite{goebel1996theoretical} \\
\hline

Cd
& 0.00000 & 47.21 & 48.25 & 47.82 & 47.22 &
47.50 $\pm$ 2.00\cite{hohm2022dipole} \\
& 0.07198 & 52.97 & \multicolumn{1}{c}{---} &
\multicolumn{1}{c}{---} & 52.75 &
54.20 $\pm$ 0.95\cite{goebel1995dispersion} \\
& 0.08383 & 55.45 & \multicolumn{1}{c}{---} &
\multicolumn{1}{c}{---} & 55.20 &
56.23 $\pm$ 0.38\cite{goebel1995dispersion} \\
& 0.14014 & 81.39 & \multicolumn{1}{c}{---} &
\multicolumn{1}{c}{---} & 80.90 &
68.80 $\pm$ 2.30\cite{goebel1995dispersion} \\
\hline

Hg
& 0.00000 & 35.13 & 35.25 & 35.24 & 35.19 &
33.92 $\pm$ 0.34\cite{goebel1996dipole} \\
& 0.07198 & 37.37 & \multicolumn{1}{c}{---} &
\multicolumn{1}{c}{---} & 37.42 &
35.75 $\pm$ 0.31\cite{goebel1996dipole} \\
& 0.08383 & 38.26 & \multicolumn{1}{c}{---} &
\multicolumn{1}{c}{---} & 38.32 &
36.63 $\pm$ 0.32\cite{goebel1996dipole} \\
& 0.14014 & 46.69 & \multicolumn{1}{c}{---} &
\multicolumn{1}{c}{---} & 46.78 &
44.64 $\pm$ 0.33\cite{goebel1996dipole} \\

\hline
\hline
\end{tabular*}
\begin{tablenotes}
\footnotesize
\item[a] FNS++4c-LR-CCSD calculation with d-aug-dyall.v4z basis set.\cite{chakraborty2025low-cost}
\item[b] X2C Hamiltonian with s-aug-dyall.v2z basis set.\cite{yuan2024formulation}
\item[c] FNS++CD-X2CMP-LR-CCSD with s-aug-dyall.v2z basis set.
\item[d] FNS++CD-X2CMP-LR-CCSD with d-aug-dyall.v4z basis set.
\end{tablenotes}
\end{threeparttable}
\end{table}

With the occupation and Cholesky thresholds fixed, and an appropriate choice of basis set and relativistic Hamiltonian, we benchmark the present FNS++CD--X2CMP--LR-CCSD implementation using the d-aug-dyall.v4z basis set for both atomic and molecular systems. The FNS++4c-LR-CCSD\cite{chakraborty2025low-cost} results in the same basis set and with the same FNS++ threshold have been used as a reference. Wherever available, the resulting polarizabilities are compared against previously reported 2c and 4c values.  Zn, Cd, and Hg are chosen as atomic benchmark systems owing to the availability of reliable experimental data for both static and dynamic polarizabilities. The static polarizability, together with three frequency-dependent polarizabilities of these atoms, evaluated at different theoretical levels, is summarized in Table~\ref{tab:zncdhg_4z} alongside the corresponding experimental values. First, we consider the X2C results reported in Ref.~\cite{yuan2024formulation} and directly compare them with our present calculations performed using the same basis set employed in that work, namely the s-aug-dyall.v2z basis set. The present results show excellent agreement with the X2C data reported in Ref.~\cite{yuan2024formulation} for static polarizability.

Systematic comparison between the FNS++4c-LR-CCSD reference polarizabilities and the corresponding FNS++CD-X2CMP-LR-CCSD results obtained with the d-aug-dyall.v4z basis set for Zn, Cd, and Hg are also presented in Table ~\ref{tab:zncdhg_4z}. This comparison provides a stringent assessment of the accuracy of the present two-component implementation. For the static polarizabilities ($\omega=0$), the FNS++CD-X2CMP-LR-CCSD values are in near-quantitative agreement with the 4c reference data for all three atoms. The deviations are typically well below 0.3~a.u., demonstrating that the scalar and spin-orbit contributions relevant to the static response are accurately captured within the X2CMP framework even when combined with the  CD treatment of two-electron integrals. 
For the dynamic polarizabilities, slightly larger deviations from the 4c results are observed at higher frequencies. Nevertheless, the mean deviation remains below 1~a.u. across all considered frequencies for Zn, Cd, and Hg, and a systematic trend can be identified. For Zn, the FNS++CD-X2CMP-LR-CCSD results consistently underestimate the corresponding 4c values, whereas for Hg, a slight overestimation is observed. Cd exhibits intermediate behavior, with deviations that remain small. 

\begin{figure}[h]
\centering
    \begin{subfigure}{0.48\textwidth} 
        \includegraphics[width=\linewidth]{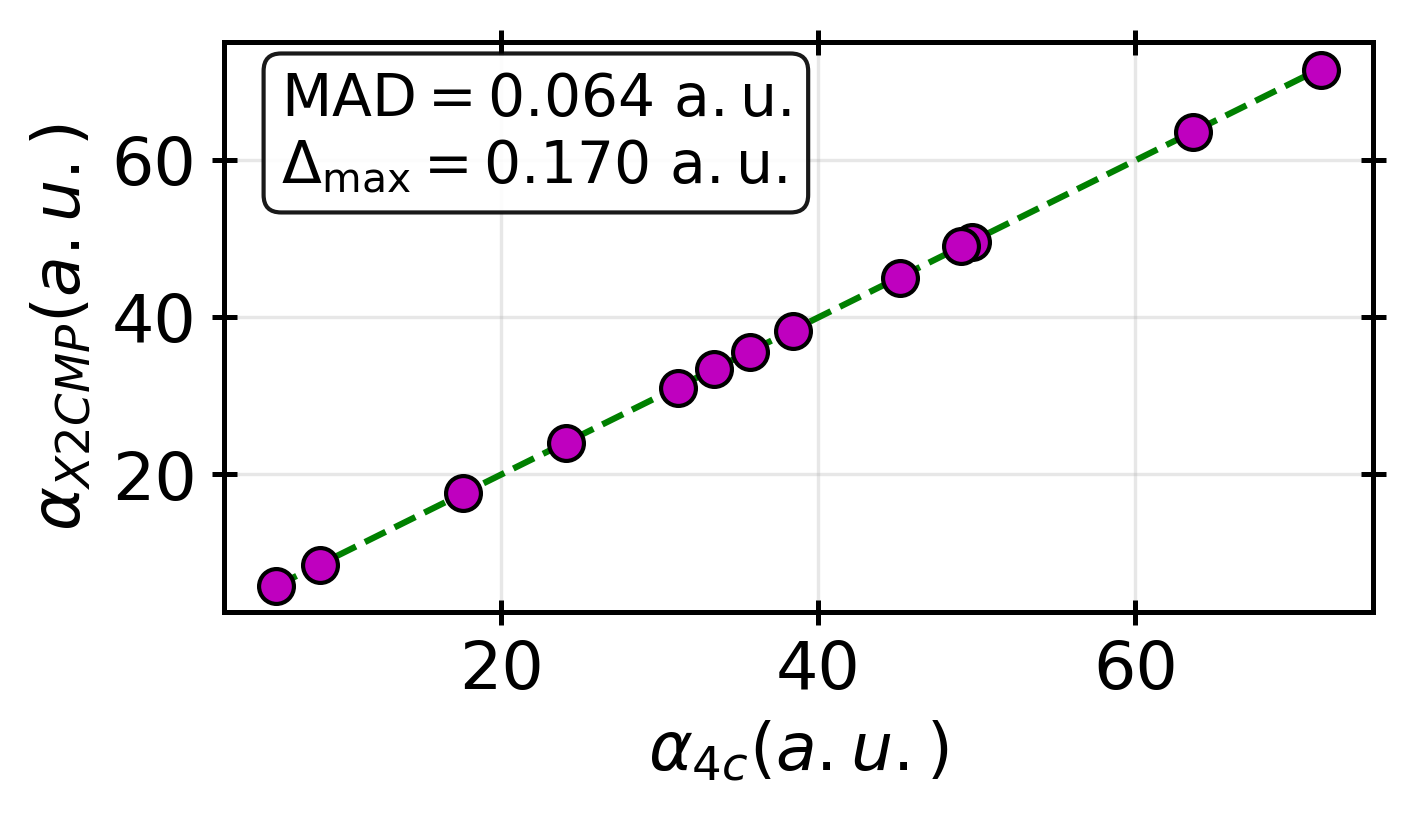} 
        \caption{}
        
    \end{subfigure}
    \begin{subfigure}{0.48\textwidth} 
        \includegraphics[width=\linewidth]{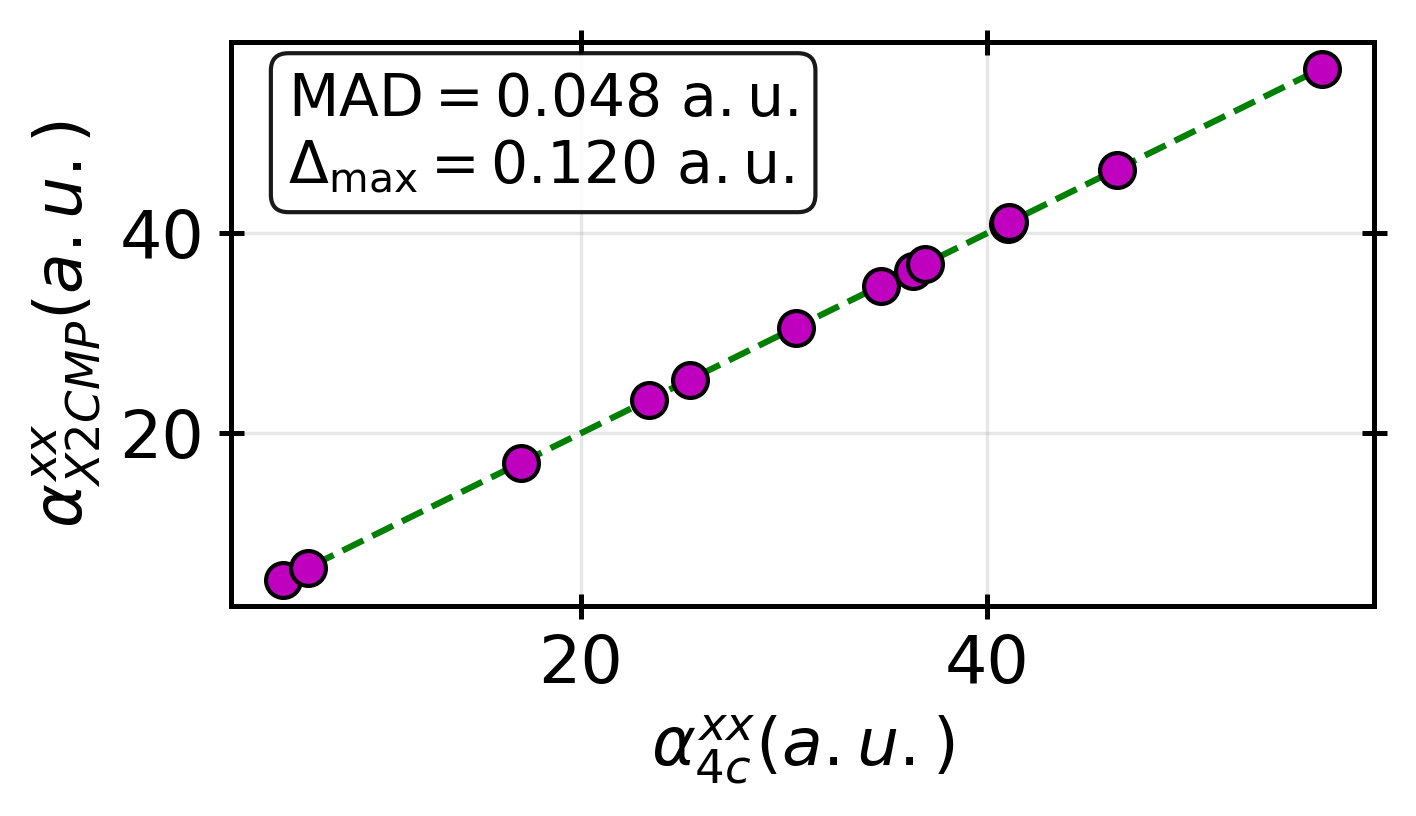} 
        \caption{}
        
    \end{subfigure}
    \begin{subfigure}{0.48\textwidth} 
        \includegraphics[width=\linewidth]{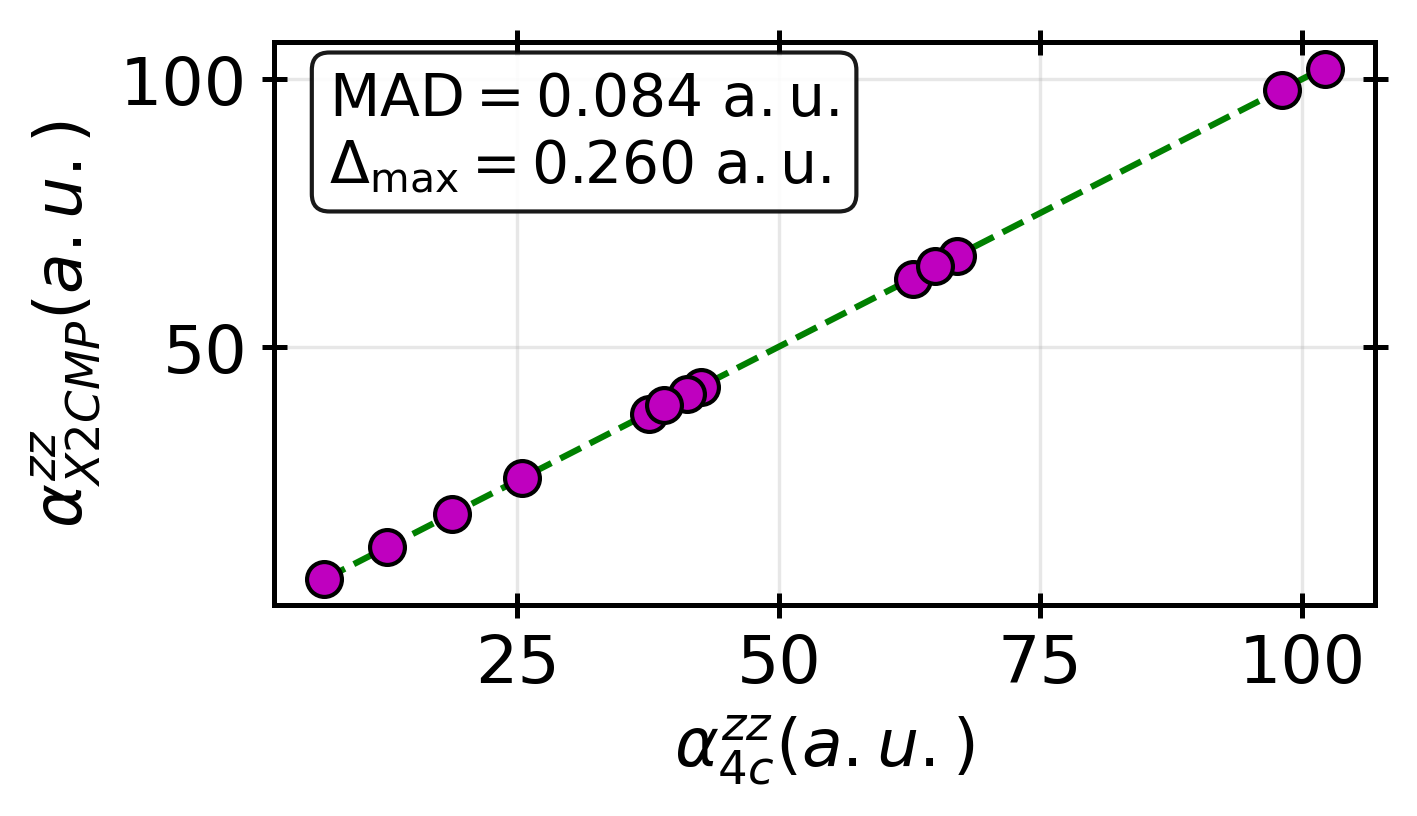} 
        \caption{}
        
    \end{subfigure}
    
    \caption{\label{fig:epsart}Correlation between (a) mean, (b) perpendicular ($\alpha_{xx}$), and (c) parallel ($\alpha_{zz}$) static polarizabilities obtained with the X2CMP Hamiltonian and four-component (4c) calculations for the set of molecules considered in Table S1.}
\end{figure}

We have also calculated the static polarizabilities of hydrogen halides (HX, X=F, Cl, Br, I), dihalogens (F$_{2}$, Cl$_{2}$, Br$_{2}$, I$_{2}$, ICl), AuH, AuF, AuCl, and HgCl$_{2}$  at the FNS++CD-X2CMP-LR-CCSD level using the d-aug-dyall.v4z basis set. The results are summarized in Table S1, alongside the corresponding FNS++4c-LR-CCSD reference values and available experimental data. The X2CMP results exhibit excellent agreement with the 4c reference values across the entire set of molecules. For both the HX and X$_{2}$ series, the perpendicular ($\alpha_{\perp}$), parallel ($\alpha_{\parallel}$), and isotropic mean($\alpha$) polarizabilities are reproduced with deviations that are typically well below $\sim$0.2 a.u. This close agreement confirms that the CD-based implementation with X2CMP Hamiltonian captures the scalar-relativistic and spin–orbit effects relevant to electric response properties, even for heavier halogens such as Br and I. 
Special attention is drawn to the heavier diatomic systems AuH, AuF, AuCl, and HgCl$_{2}$, which represent particularly demanding test cases due to their large number of electrons and strong relativistic effects. For these systems, the FNS++CD-X2CMP-LR-CCSD results reproduce the corresponding 4c polarizabilities almost quantitatively, with deviations remaining well below 0.1~a.u. for the isotropic mean values. This level of agreement demonstrates that the X2CMP Hamiltonian retains the essential relativistic effects governing the electric response even in the presence of heavy nuclei and significant spin-orbit coupling.

The quality of the agreement is further illustrated in Fig.~\ref{fig:epsart}, which shows the correlation between mean static polarizabilities obtained from X2CMP and 4c calculations for the diatomic systems listed in Table S1. The near-perfect linear correlation demonstrates that the X2CMP results track the 4c reference values almost quantitatively over a wide polarizability range (approximately 5-70~a.u.). The small mean absolute deviation (MAD = 0.064~a.u.) and maximum deviation ($\Delta_{\text{max}}$ = 0.170 a.u.) demonstrate the numerical precision of the present implementation. 
\begin{table}
\caption{\label{tab:table1}
Dynamic polarizabilities (a.u.) of I$_2$ obtained using a $10^{-5}$ FNS++ occupation threshold.
}

\centering
\begin{threeparttable}

\begin{tabular*}{\textwidth}{@{\extracolsep{\fill}}lccc@{}}
\hline
\hline
Method & $\alpha_{\perp}$ & $\alpha_{\parallel}$ & $\alpha$ \\
\hline

\multicolumn{4}{c}{\textbf{$\omega = 0.07198$ a.u.}}\\
\hline
X2C-HF\tnote{a}      & 55.0 & 152.0 & 87.4 \\
X2C-B3LYP\tnote{a}   & 58.7 & -10.7 & 35.6 \\
X2C-CC\tnote{a}      & 55.8 & 114.8 & 75.5 \\
4c-CC\tnote{b}       & 56.8 & 115.9 & 76.5 \\
X2CMP-CC\tnote{c}    & 56.8 & 115.9 & 76.5 \\
4c-CC\tnote{d}       & 57.2 & 114.2 & 76.2 \\
X2CMP-CC\tnote{e}    & 56.3 & 113.6 & 75.4 \\
Expt.\cite{maroulis1997electrooptical}
                     & --- & --- & $86.8 \pm 2.2$ \\

\hline

\multicolumn{4}{c}{\textbf{$\omega = 0.07669$ a.u.}}\\
\hline
X2C-HF\tnote{a}      & 56.0 & -97.3 & 4.9 \\
X2C-B3LYP\tnote{a}   & 62.0 & 75.4 & 66.5 \\
X2C-CC\tnote{a}      & 56.8 & 124.0 & 79.2 \\
4c-CC\tnote{b}       & 57.7 & 125.4 & 80.3 \\
X2CMP-CC\tnote{c}    & 57.7 & 125.9 & 80.4 \\
4c-CC\tnote{d}       & 58.4 & 119.1 & 78.7 \\
X2CMP-CC\tnote{e}    & 58.2 & 118.0 & 78.1 \\
Expt.\cite{maroulis1997electrooptical}
                     & --- & --- & $93.6 \pm 3.4$ \\

\hline

\multicolumn{4}{c}{\textbf{$\omega = 0.14014$ a.u.}}\\
\hline
X2C-HF\tnote{a}      & 55.3 & 117.9 & 76.2 \\
X2C-B3LYP\tnote{a}   & 61.0 & 114.5 & 78.8 \\
X2C-CC\tnote{a}      & 59.9 & 113.9 & 77.9 \\
4c-CC\tnote{b}       & 60.9 & 114.7 & 78.8 \\
X2CMP-CC\tnote{c}    & 60.8 & 114.6 & 78.8 \\
4c-CC\tnote{d}       & 63.0 & 117.6 & 81.2 \\
X2CMP-CC\tnote{e}    & 62.9 & 117.4 & 81.1 \\
Expt.\cite{maroulis1997electrooptical}
                     & --- & --- & $95.3 \pm 1.9$ \\

\hline
\hline
\end{tabular*}

\begin{tablenotes}[flushleft]
\footnotesize
\item[a] X2C Hamiltonian with the s-aug-dyall.v2z basis set. Values are taken from Ref.~\cite{yuan2024formulation}.
\item[b] FNS++4c-LRCCSD with the s-aug-dyall.v2z basis set.
\item[c] FNS++CD-X2CMP-LRCCSD with the s-aug-dyall.v2z basis set.
\item[d] FNS++4c-LRCCSD with the d-aug-dyall.v4z basis set.
\item[e] FNS++CD-X2CMP-LRCCSD with the d-aug-dyall.v4z basis set.
\end{tablenotes}

\end{threeparttable}
\end{table}

Table~\ref{tab:table1} summarizes the dynamic polarizabilities of I$_2$ computed at three frequencies, $ 0.07198$, $0.07669$, and $0.14014$ a.u.,  allowing a systematic assessment of the performance of our X2CMP implementation against 4c results and available experimental data. Across all three frequencies, the X2CMP-CC results obtained with the smaller s-aug-dyall.v2z basis set are in excellent agreement with the corresponding 4c-CC values, with deviations in the mean polarizability $\alpha$ typically within 0.1--0.2 a.u. This close correspondence validates the fidelity of the X2CMP approach in reproducing four-component relativistic effects at a reduced computational cost. Notably, for $\omega = 0.07198$ a.u., both the 4c-CC and X2CMP-CC methods yield $\alpha_{\perp} = 56.8$ a.u., $\alpha_{\parallel} = 115.9$ a.u., and $\bar{\alpha} = 76.5$ a.u. with the smaller basis, demonstrating near-perfect agreement between the two formalisms. Upon going to the larger d-aug-dyall.v4z basis set, both the 4c-CC and X2CMP-CC results shift modestly but consistently. For instance, at $\omega = 0.07198$ a.u., the mean polarizability decreases slightly from 76.5 to 76.2 (4c-CC) and from 76.5 to 75.4 (X2CMP-CC), indicating a mild basis set sensitivity. The X2CMP-CC values with the larger basis remain in close agreement with the 4c-CC counterparts, with differences not exceeding 1.0 a.u. in $\bar{\alpha}$ across all frequencies considered.
It is worth noting the anomalous behavior exhibited by the X2C-HF and X2C-B3LYP methods at $\omega = 0.07669$ a.u., where $\alpha_{\parallel}$ shows large deviating values of $-97.3$ and $75.4$ a.u., respectively. The strongly negative $\alpha_{\parallel}$ for X2C-HF near this frequency is a signature of a near-resonance condition, wherein the applied frequency approaches an electronic excitation energy of the molecule, causing a divergence in the linear response. The coupled-cluster methods, by contrast, exhibit physically consistent polarizabilities across all three frequencies, underscoring the importance of an adequate treatment of electron correlation for reliable dynamic polarizability predictions in heavy-element systems such as I$_2$. Compared with the experimental values of Maroulis~,\cite{maroulis1997electrooptical} the CC-level methods systematically underestimate $\bar{\alpha}$ at all three frequencies. At $\omega = 0.07198$ a.u., the experimental value is $86.8 \pm 2.2$ a.u., while the best X2CMP-CC result (with d-aug-dyall.v4z) yields $75.4$ a.u., a discrepancy of approximately 11 a.u. A similar systematic gap is observed at the other two frequencies. This underestimation may be attributed to the truncation of the cluster operator at the singles and doubles level, or to the neglect of vibrational effects in the calculations. Further investigation of these effects is outside the scope of the present study.

\subsection{Computational efficiency}
\begin{figure}[h]
\centering
    \begin{subfigure}{0.7\textwidth} 
        \includegraphics[width=\linewidth]{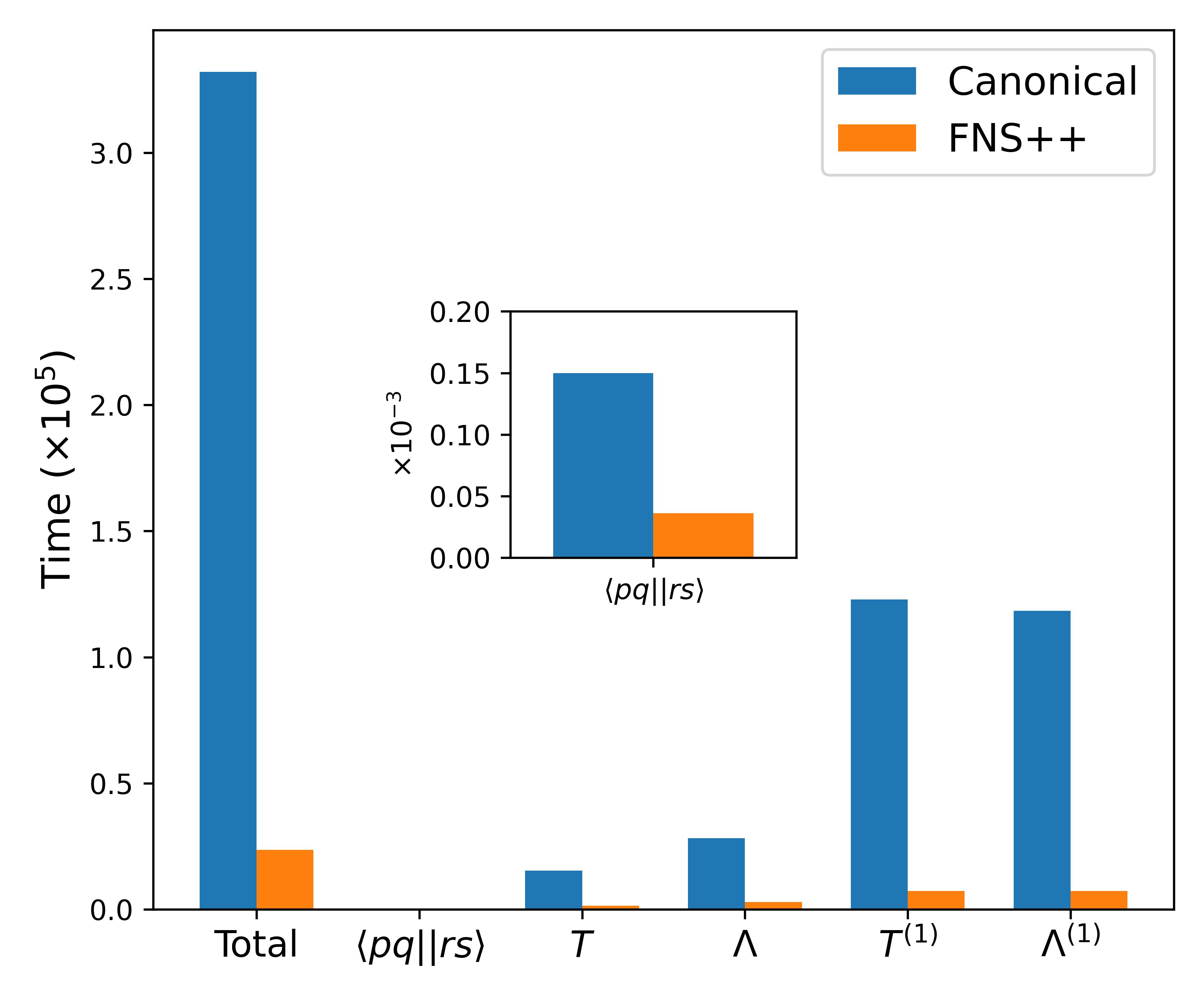} 
    \end{subfigure}
    
    \caption{\label{fig:auf_time}Wall times (in seconds) of the CD-LRCCSD calculation on the AuF molecule using the s-aug-dyall.v2z (for Au) and unc-aug-cc-pVDZ (for F) basis sets.}
\end{figure}

The computational efficiency of the present FNS++ scheme relative to the canonical approach is illustrated in Fig.~\ref{fig:auf_time}, which presents the wall times for various sections of the FNS++CD-X2CMP-LRCCSD calculation on the AuF molecule using the s-aug-dyall.v2z ( for Au) and unc-aug-cc-pVDZ (for F) basis sets, with the virtual spinor space truncated at 825 a.u., retaining 300 virtual spinors in the canonical space. The calculation has been performed on a 16-core Intel Xeon Gold 5315Y CPU @ 3.20 GHz system equipped with 512 GB of RAM.

The most striking observation is the dramatic reduction in the total wall time afforded by the FNS++ scheme. The canonical calculation requires approximately 3 days, 20 hours, 15 mins, whereas the FNS++ calculation completes in roughly 6 hours, 33 mins, a speedup of nearly 14-fold. This substantial gain demonstrates the practical advantage of the FNS++ truncation strategy when combined with the CD-based approximation. Examining the individual computational sections reveals the origin of these savings. The two-electron integral transformation step, denoted $\langle pq \| rs \rangle$, is one of the primary bottlenecks in any correlated wavefunction calculation. As highlighted in the inset of Fig.~\ref{fig:auf_time}, the integral transformation time is negligible in the present case owing to the use of the CD approximation. The amplitude equations for $T$ and $\Lambda$ exhibit notable speedups, of nearly 10--11 times compared to the canonical calculation. Perhaps the most significant individual contributions to the overall speedup come from the response amplitude equations, $T^{(1)}$ and $\Lambda^{(1)}$, which govern the linear response of the coupled-cluster ground-state wave function to an external perturbation. In the canonical framework, $T^{(1)}$ and $\Lambda^{(1)}$ each require approximately 1 day 10 hours, collectively accounting for the dominant portion of the total wall time. Under the FNS++CD scheme, both contributions are reduced to below 2.5 hours, resulting in a speedup exceeding 14-fold for each section.

\begin{figure}[h]
\centering
    \begin{subfigure}{0.4\textwidth} 
        \includegraphics[width=\linewidth]{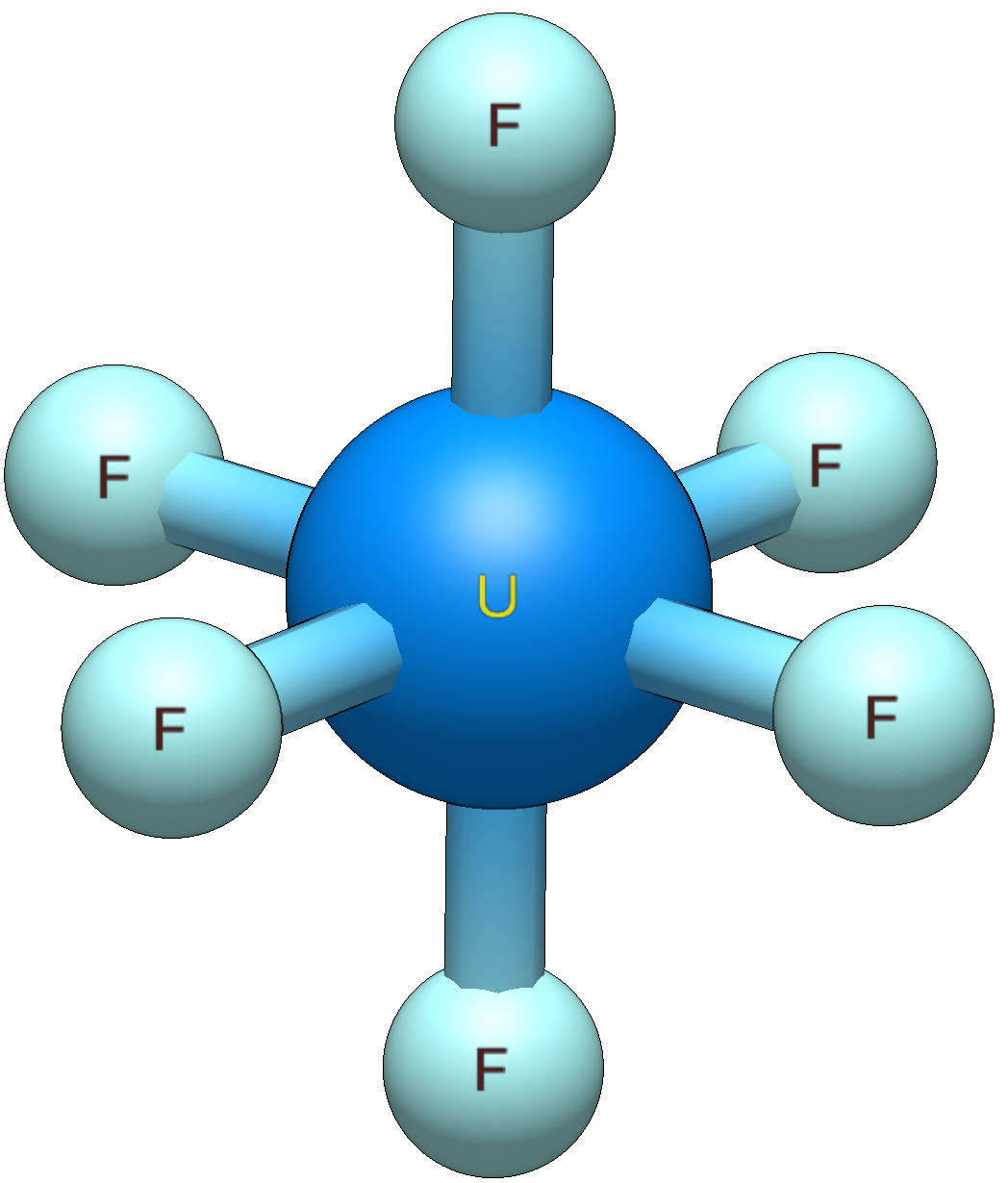} 
        \caption{}
        
    \end{subfigure}
    
    \caption{\label{fig:uf6} Uranium Hexafluoride (UF$_{6}$) complex.}
\end{figure}

To demonstrate the capability of the present implementation for treating large and non-linear systems, we consider the UF$_6$ complex as a representative example (see Fig.~\ref{fig:uf6}). The static polarizability of this system has been computed using a triple-$\zeta$ basis set, where the s-aug-dyall.v3z basis is employed for uranium, and an uncontracted aug-cc-pVTZ basis for fluorine. The calculation was carried out on a dedicated workstation equipped with dual Intel(R) Xeon(R) Gold 5315Y processors (3.20 GHz) and 2.0 TB of RAM. The system comprises 146 electrons and 1338 virtual spinors in the canonical space. Upon applying the frozen-core approximation along with an FNS++ truncation threshold of $10^{-5}$, the active space is reduced to 66 occupied and 388 virtual spinors. 
The total wall time for the calculation amounts to 6 days, 18 hours, and 48 minutes. Notably, the integral transformation step is highly efficient, requiring only 85 seconds. The ground-state coupled-cluster and lambda amplitude equations require 3 days, 10 hours, and 55 minutes, whereas the response amplitude equations require 1 day, 8 hours, and 8 minutes.
The computed static polarizability is 55.8 a.u., which is in excellent agreement with the experimental value of $54.4 \pm 7.0$ a.u.~\cite{hohm2013experimental} This clearly demonstrates that the present approach achieves high accuracy while remaining computationally feasible for large-scale relativistic systems.

Since the present implementation enables the computation of frequency-dependent properties, we have also evaluated the dynamic polarizability of the UF$_6$ molecule at an external frequency of 0.07198 a.u., employing the same basis sets and computational setup as used for the static case. The number of virtual spinors after truncation increases slightly to 392, which is expected, as the perturbed amplitudes entering the construction of the FNS++ densities are explicitly frequency dependent. The total computational time amounts to 7 days, 2 hours, and 41 minutes. Of this, the evaluation of the ground-state left and right amplitudes required 3 days, 21 hours, and 6 minutes, while the response calculation took 1 day, 4 hours, and 17 minutes. The computed polarizability at this frequency is 57.16 a.u. To the best of our knowledge, no experimental reference value is available for UF$_6$ at this frequency. Compared to the static limit (0.0 a.u.), the polarizability increases by 1.36 a.u. This behavior is consistent with the general trend of dynamic polarizability, which typically exhibits a gradual increase with frequency when the applied field is sufficiently far from any electronic excitation resonance.

\begin{figure}[h]
\centering
    \begin{subfigure}{0.6\textwidth} 
        \includegraphics[width=\linewidth]{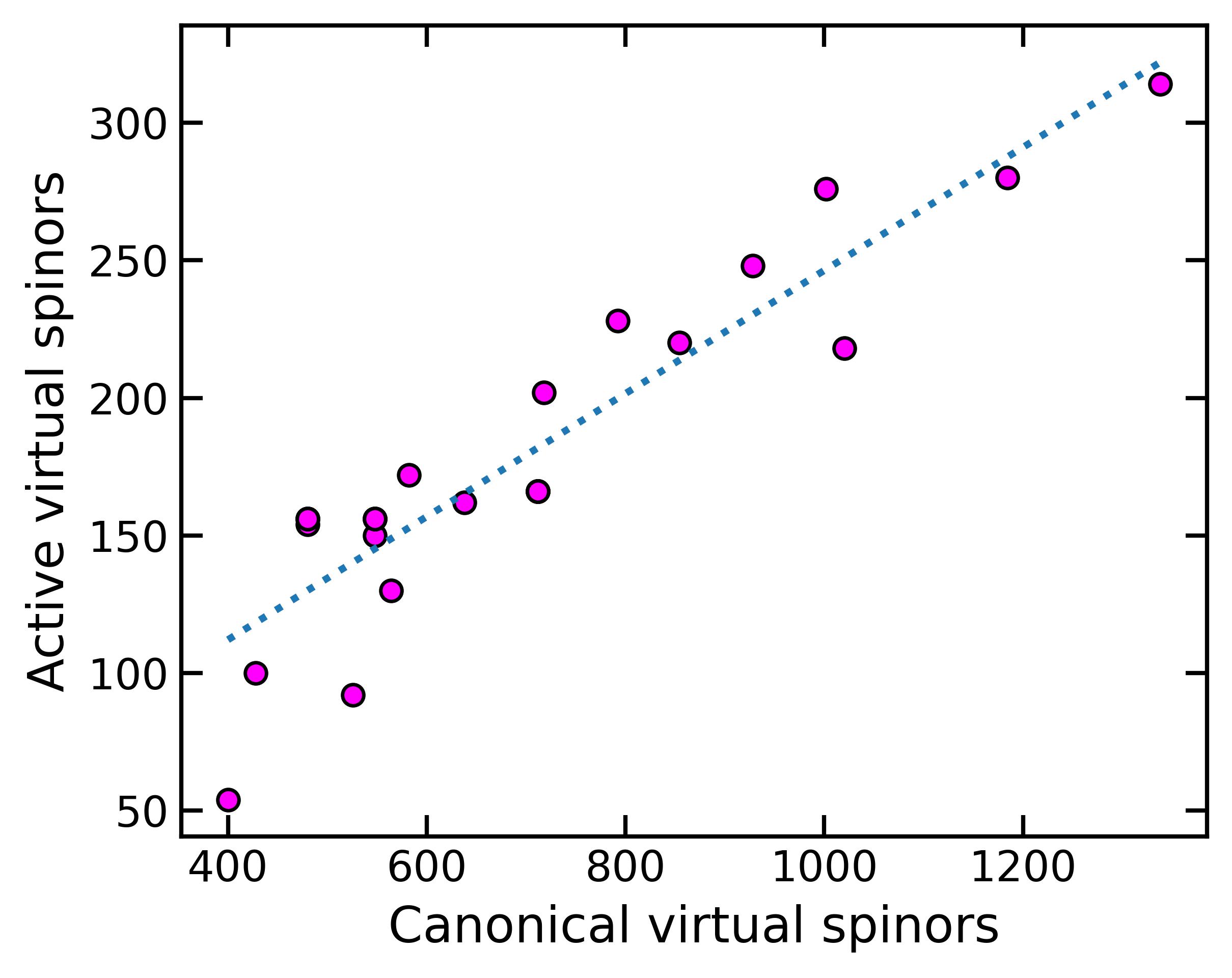} 
        %
    \end{subfigure}
    
    \caption{\label{fig:scatter_fns_vs_can} Correlation between the number of canonical and active virtual spinors after truncation obtained using the FNS++ scheme.}
\end{figure}

As a final remark, we examine the extent of virtual space truncation achieved across the systems considered in this study. Figure~\ref{fig:scatter_fns_vs_can} illustrates the correlation between the number of canonical and active virtual spinors after truncation. As evident from the figure, the canonical virtual space spans a wide range, approximately from 400 to 1300 spinors, depending on the system size. In contrast, upon applying the FNS++ truncation with a threshold of $10^{-5}$, the virtual space is significantly reduced to a much narrower range of about 50--300 spinors. This substantial reduction highlights the effectiveness of the truncation scheme in systematically compressing the virtual space. On average, nearly 73\% of the virtual spinors are removed, demonstrating a consistent and robust performance across different systems. 

\section{Conclusion}

In this work, we present an efficient implementation of a low-cost linear-response coupled-cluster singles and doubles (LR-CCSD) method based on the X2CAMF and X2CMP Hamiltonians for the calculation of static and frequency-dependent polarizabilities in systems exhibiting both relativistic effects and significant electron correlation. Building upon our earlier 4c-FNS++-LR-CCSD framework, the present approach employs the X2C-based Hamiltonians for reference state construction, along with Cholesky decomposition (CD) to significantly reduce memory requirements. In the current work, the storage of costly three- and four-index integrals is avoided by generating the required integrals and intermediates on the fly. Benchmark investigations reveal that, particularly for large and highly augmented basis sets, the X2CAMF scheme may lead to numerical instabilities, which can be avoided by using the X2CMP Hamiltonian. The FNS++CD-X2CMP-LR-CCSD method demonstrates excellent agreement with the four-component reference results across a wide range of frequencies and molecular systems, confirming its accuracy and efficiency. Two approaches for constructing the FNS++ basis have been examined: one based on a single averaged density obtained from the three Cartesian directions, and another using direction-specific densities for each individual Cartesian component. It is observed that the direction-specific approach provides improved accuracy when a very small number of virtual orbitals is retained after truncation. However, this improvement comes at a significantly higher computational cost, as it requires repeated ground-state coupled cluster calculations and integral transformations for each direction. In contrast, the averaged density approach avoids these redundancies while still delivering reliable results, and is therefore adopted in the present work as a more computationally efficient strategy. 

Calculations performed using the averaged density, in combination with Cholesky decomposition (CD) and the X2CMP Hamiltonian, yield a mean absolute deviation (MAD) of 0.064 a.u. in the mean polarizability across a diverse set of molecules when compared to four-component results. Furthermore, a comparison between the canonical and the present FNS++ based approaches demonstrates an approximate 14-fold computational speedup when employing the CD framework with the X2CMP Hamiltonian, and it is also important to note that the present implementation significantly reduces memory requirements. To illustrate the applicability of the method to general non-linear systems with strong relativistic effects, we have computed the static and dynamic polarizabilities of the UF$_6$ molecule, obtaining a result in good agreement with the experimental value. 
The present implementation does not contain the orbital relaxation effect, which can be significant in some cases. Additionally, the incorporation of higher-order electron correlation is expected to further improve the accuracy of coupled-cluster response properties. The inclusion of triple excitation in the relativistic LR-CC framework can be one of the potential remedies to these problems. Work in this direction is currently in progress.

\section{Acknowledgments}
AKD, SC, and MB acknowledge financial support from IIT Bombay, the ANRF (CRG/2023/002558), and ISRO. The authors also acknowledge the IIT Bombay supercomputing facility and C-DAC (Param Smriti, Param Brahma, and Param Rudra) for providing computational resources. SC acknowledges the Prime Minister's Research Fellowship (PMRF).

\bibliography{references}

@misc{socutils,
  author       = {Xubwa},
  title        = {socutils},
  year         = {2025},
  url          = {https://github.com/xubwa/socutils},
  note         = {Accessed: 2025-07-30}
}

@incollection{Champagne,
    author = {Champagne, Benoît},
    editor = {Springborg, Michael},
    isbn = {978-1-84755-881-7},
    title = {Polarizabilities and hyperpolarizabilities },
    booktitle = {Chemical Modelling},
    publisher = {Royal Society of Chemistry},
    year = {2009},
    month = {09},
    doi = {10.1039/b812904p},
    url = {https://doi.org/10.1039/b812904p},
}

@article{datta1995coupled,
author = {Datta, B. and Sen, P. and Mukherjee, D.},
title = {Coupled-Cluster Based Linear Response Approach to Property Calculations: Dynamic Polarizability and Its Static Limit},
journal = {The Journal of Physical Chemistry},
volume = {99},
number = {17},
pages = {6441-6451},
year = {1995},
doi = {10.1021/j100017a024},
URL = {https://doi.org/10.1021/j100017a024}
}

@article{Hendrik77,
  title={Calculation of properties with the coupled-cluster method},
  author={Monkhorst, Hendrik J},
  journal={Int. J. Quantum Chem. },
  volume={12},
  number={S11},
  pages={421--432},
  year={1977},
  publisher={Wiley Online Library},
  URL= {https://onlinelibrary.wiley.com/doi/abs/10.1002/qua.560120850}
}

@article{Helgaker1990,
    author = {Koch, Henrik and Jensen, Hans Jo/rgen Aa. and Jo/rgensen, Poul and Helgaker, Trygve},
    title = {Excitation energies from the coupled cluster singles and doubles linear response function (\uppercase{CCSDLR}). Applications to \uppercase{B}e, \uppercase{CH+}, \uppercase{CO}, and \uppercase{H2O}},
    journal = {J. Chem. Phys. },
    volume = {93},
    number = {5},
    pages = {3345-3350},
    year = {1990},
    month = {09},
    issn = {0021-9606},
    doi = {10.1063/1.458815},
    url = {https://doi.org/10.1063/1.458815},
}

@article{sekino1984linear,
  title={A linear response, coupled-cluster theory for excitation energy},
  author={Sekino, Hideo and Bartlett, Rodney J},
  journal={Int. J. Quantum Chem. },
  volume={26},
  number={S18},
  pages={255--265},
  year={1984},
  publisher={Wiley Online Library},
  URL= {https://onlinelibrary.wiley.com/doi/10.1002/qua.560260826} 
}

@incollection{casida1995time,
  title={Time-dependent density functional response theory for molecules},
  author={Casida, Mark E},
  booktitle={Recent Advances In Density Functional Methods: (Part I)},
  pages={155--192},
  year={1995},
  publisher={World Scientific},
  URL={https://www.worldscientific.com/doi/abs/10.1142/9789812830586_0005}}

@article{christiansen1995second,
  title={The second-order approximate coupled cluster singles and doubles model \uppercase{CC2}},
  author={Christiansen, Ove and Koch, Henrik and J{\o}rgensen, Poul},
  journal={Chemical Physics Letters},
  volume={243},
  number={5-6},
  pages={409--418},
  year={1995},
  publisher={Elsevier},
  url = {https://www.sciencedirect.com/science/article/pii/000926149500841Q}
}

@article{christiansen1998response,
  title={Response functions from Fourier component variational perturbation theory applied to a time-averaged quasienergy},
  author={Christiansen, Ove and J{\o}rgensen, Poul and H{\"a}ttig, Christof},
  journal={Int. J. Quantum Chem.},
  volume={68},
  number={1},
  pages={1--52},
  year={1998},
  publisher={Wiley Online Library},
  url = {https://onlinelibrary.wiley.com/doi/abs/10.1002/%28SICI%291097-461X%281998%2968%3A1%3C1%3A%3AAID-QUA1%3E3.0.CO%3B2-Z}
}

@article{christiansen1999frequency,
  title={Frequency-dependent polarizabilities and first hyperpolarizabilities of \uppercase{CO} and \uppercase{H2O} from coupled cluster calculations},
  author={Christiansen, Ove and Gauss, J{\"u}rgen and Stanton, John F},
  journal={Chemical physics letters},
  volume={305},
  number={1-2},
  pages={147--155},
  year={1999},
  publisher={Elsevier},
  url = {https://www.sciencedirect.com/science/article/pii/S0009261499003589}
}

@article{hald2003calculation,
  title={Calculation of frequency-dependent polarizabilities using the approximate coupled-cluster triples model \uppercase{CC3}},
  author={Hald, Kasper and Paw{\l}owski, Filip and J{\o}rgensen, Poul and H{\"a}ttig, Christof},
  journal={J. Chem. Phys. },
  volume={118},
  number={3},
  pages={1292--1300},
  year={2003},
  publisher={American Institute of Physics},
  URL = {https://pubs.aip.org/aip/jcp/article/118/3/1292/463640/Calculation-of-frequency-dependent}
}

@article{kobayashi1994calculation,
  title={Calculation of frequency-dependent polarizabilities using coupled-cluster response theory},
  author={Kobayashi, Rika and Koch, Henrik and J{\o}rgensen, Poul},
  journal={Chemical physics letters},
  volume={219},
  number={1-2},
  pages={30--35},
  year={1994},
  publisher={Elsevier},
  URL = {https://www.sciencedirect.com/science/article/pii/0009261494000514}
}

@article{nielsen1980transition,
  title={Transition moments and dynamic polarizabilities in a second order polarization propagator approach},
  author={Nielsen, Egon S and Jo/rgensen, Poul and Oddershede, Jens},
  journal={J. Chem. Phys. },
  volume={73},
  number={12},
  pages={6238--6246},
  year={1980},
  publisher={American Institute of Physics}, 
  URL= {https://pubs.aip.org/aip/jcp/article/73/12/6238/451114/Transition-moments-and-dynamic-polarizabilities-in} 
}

@article{hammond2008coupled,
  title={Coupled-cluster dynamic polarizabilities including triple excitations},
  author={Hammond, Jeff R and De Jong, Wibe A and Kowalski, Karol},
  journal={J. Chem. Phys. },
  volume={128},
  number={22},
  year={2008},
  publisher={AIP Publishing},
  URL = {https://pubs.aip.org/aip/jcp/article/128/22/224102/924805/Coupled-cluster-dynamic-polarizabilities-including}
}

@article{hammond2009accurate,
  title={Accurate dipole polarizabilities for water clusters n= 2--12 at the coupled-cluster level of theory and benchmarking of various density functionals},
  author={Hammond, Jeff R and Govind, Niranjan and Kowalski, Karol and Autschbach, Jochen and Xantheas, Sotiris S},
  journal={J. Chem. Phys. },
  volume={131},
  number={21},
  year={2009},
  publisher={AIP Publishing},
  URL={https://pubs.aip.org/aip/jcp/article/131/21/214103/955011/Accurate-dipole-polarizabilities-for-water} 
}

@article{saue2003linear,
  title={Linear response at the 4-component relativistic level: Application to the frequency-dependent dipole polarizabilities of the coinage metal dimers},
  author={Saue, Trond and Jensen, HJ Aa},
  journal={J. Chem. Phys. },
  volume={118},
  number={2},
  pages={522--536},
  year={2003},
  publisher={American Institute of Physics},
  URL={https://pubs.aip.org/aip/jcp/article/118/2/522/460855/Linear-response-at-the-4-component-relativistic}
}

@article{salek2005linear,
  title={Linear response at the 4-component relativistic density-functional level: application to the frequency-dependent dipole polarizability of Hg, AuH and PtH2},
  author={Salek, Pawel and Helgaker, Trygve and Saue, Trond},
  journal={Chemical physics},
  volume={311},
  number={1-2},
  pages={187--201},
  year={2005},
  publisher={Elsevier},
  url={https://www.sciencedirect.com/science/article/pii/S0301010404005749}
}

@article{hait2018accurate,
  title={How accurate are static polarizability predictions from density functional theory? An assessment over 132 species at equilibrium geometry},
  author={Hait, Diptarka and Head-Gordon, Martin},
  journal={Phys. Chem. Chem. Phys.,},
  volume={20},
  number={30},
  pages={19800--19810},
  year={2018},
  publisher={Royal Society of Chemistry},
  url={https://pubs.rsc.org/en/content/articlelanding/2018/cp/c8cp03569e}
}

@article{burke2012perspective,
  title={Perspective on density functional theory},
  author={Burke, Kieron},
  journal={J. Chem. Phys. },
  volume={136},
  number={15},
  year={2012},
  publisher={AIP Publishing},
  url={https://pubs.aip.org/aip/jcp/article/136/15/150901/941589/Perspective-on-density-functional-theory}
}

@article{salek2005comparison,
  title={A comparison of density-functional-theory and coupled-cluster frequency-dependent polarizabilities and hyperpolarizabilities},
  author={Sa{\l}ek, Pawe{\l} and Helgaker, Trygve and Vahtras, Olav and {\AA}gren, Hans and Jonsson, Dan and Gauss, J{\"u}rgen},
  journal={Molecular Physics},
  volume={103},
  number={2-3},
  pages={439--450},
  year={2005},
  publisher={Taylor \& Francis},
  url={https://www.tandfonline.com/doi/full/10.1080/00268970412331319254}
}

@article{gauss1998triple,
  title={Triple excitation effects in coupled-cluster calculations of frequency-dependent hyperpolarizabilities},
  author={Gauss, J{\"u}rgen and Christiansen, Ove and Stanton, John F},
  journal={Chemical physics letters},
  volume={296},
  number={1-2},
  pages={117--124},
  year={1998},
  publisher={Elsevier},
  url={https://www.sciencedirect.com/science/article/pii/S0009261498010136}
}

@article{larsen1999polarizabilities,
  title={Polarizabilities and first hyperpolarizabilities of HF, Ne, and BH from full configuration interaction and coupled cluster calculations},
  author={Larsen, Helena and Olsen, Jeppe and H{\"a}ttig, Christof and Jo/rgensen, Poul and Christiansen, Ove and Gauss, J{\"u}rgen},
  journal={J. Chem. Phys. },
  volume={111},
  number={5},
  pages={1917--1925},
  year={1999},
  publisher={American Institute of Physics},
  url={https://pubs.aip.org/aip/jcp/article/111/5/1917/531621/Polarizabilities-and-first-hyperpolarizabilities}
}

@article{pyykko1988relativistic,
  title={Relativistic effects in structural chemistry},
  author={Pyykko, Pekka},
  journal={Chemical Reviews},
  volume={88},
  number={3},
  pages={563--594},
  year={1988},
  publisher={ACS Publications},
  url = {https://doi.org/10.1021/cr00085a006}
}

@article{swirles1935relativistic,
  title={The relativistic self-consistent field},
  author={Swirles, Bertha},
  journal={Proceedings of the Royal Society of London. Series A-Mathematical and Physical Sciences},
  volume={152},
  number={877},
  pages={625--649},
  year={1935},
  publisher={The Royal Society London},
  URL = {https://royalsocietypublishing.org/doi/abs/10.1098/rspa.1935.0211},
}

@book{dyall2007introduction,
    author = {Dyall, Kenneth G and Faegri, Knut},
    title = {Introduction to Relativistic Quantum Chemistry},
    publisher = {Oxford University Press},
    year = {2007},
    month = {07},
    isbn = {9780195140866},
    doi = {10.1093/oso/9780195140866.001.0001},
    url = {https://doi.org/10.1093/oso/9780195140866.003.0005},
}

@article{vcivzek1966correlation,
  title={On the correlation problem in atomic and molecular systems. Calculation of wavefunction components in Ursell-type expansion using quantum-field theoretical methods},
  author={{\v{C}}{\'\i}{\v{z}}ek, Ji{\v{r}}{\'\i}},
  journal={J. Chem. Phys. },
  volume={45},
  number={11},
  pages={4256--4266},
  year={1966},
  publisher={AIP Publishing},
  url={https://pubs.aip.org/aip/jcp/article/45/11/4256/211247/On-the-Correlation-Problem-in-Atomic-and-Molecular}
}

@article{vcivzek1969use,
  title={On the use of the cluster expansion and the technique of diagrams in calculations of correlation effects in atoms and molecules},
  author={{\v{C}}{\'\i}{\v{z}}ek, Ji{\v{r}}{\'\i}},
  journal={Advances in chemical physics},
  volume={14},
  pages={35--89},
  year={1969},
  url={https://onlinelibrary.wiley.com/doi/10.1002/9780470143599.ch2}
}

@article{vcivzek1991origins,
  title={Origins of coupled cluster technique for atoms and molecules},
  author={{\v{C}}{\'\i}{\v{z}}ek, Ji{\v{r}}{\'\i}},
  journal={Theoretica chimica acta},
  volume={80},
  pages={91--94},
  year={1991},
  publisher={Springer},
  url={https://link.springer.com/article/10.1007/BF01119616}
}

@incollection{paldus2005beginnings,
  title={The beginnings of coupled-cluster theory: An eyewitness account},
  author={Paldus, Josef},
  booktitle={Theory and Applications of Computational Chemistry},
  pages={115--147},
  year={2005},
  publisher={Elsevier},
  url = {https://www.sciencedirect.com/science/article/pii/B9780444517197500500},
}

@article{crawford2007introduction,
  title={An introduction to coupled cluster theory for computational chemists},
  author={Crawford, T Daniel and Schaefer III, Henry F},
  journal={Reviews in computational chemistry},
  volume={14},
  pages={33--136},
  year={2007},
  url={https://onlinelibrary.wiley.com/doi/10.1002/9780470125915.ch2}
}

@article{chamoli2022reduced,
  title={A reduced cost four-component relativistic coupled cluster method based on natural spinors},
  author={Chamoli, Somesh and Surjuse, Kshitijkumar and Jangid, Bhavnesh and Nayak, Malaya K and Dutta, Achintya Kumar},
  journal={J. Chem. Phys. },
  volume={156},
  number={20},
  year={2022},
  publisher={AIP Publishing},
  url={https://pubs.aip.org/aip/jcp/article/156/20/204120/2841325/A-reduced-cost-four-component-relativistic-coupled}
}

@article{surjuse2022low,
  title={A low-cost four-component relativistic equation of motion coupled cluster method based on frozen natural spinors: Theory, implementation, and benchmark},
  author={Surjuse, Kshitijkumar and Chamoli, Somesh and Nayak, Malaya K and Dutta, Achintya Kumar},
  journal={J. Chem. Phys. },
  volume={157},
  number={20},
  year={2022},
  publisher={AIP Publishing},
  url={https://pubs.aip.org/aip/jcp/article/157/20/204106/2842109/A-low-cost-four-component-relativistic-equation-of}
}

@article{yuan2022assessing,
  title={Assessing MP2 frozen natural orbitals in relativistic correlated electronic structure calculations},
  author={Yuan, Xiang and Visscher, Lucas and Gomes, Andr{\'e} Severo Pereira},
  journal={J. Chem. Phys. },
  volume={156},
  number={22},
  year={2022},
  publisher={AIP Publishing},
  url={https://pubs.aip.org/aip/jcp/article/156/22/224108/2841380/Assessing-MP2-frozen-natural-orbitals-in}
}

@article{majee2024reduced,
    author = {Majee, Kamal and Chakraborty, Sudipta and Mukhopadhyay, Tamoghna and Nayak, Malaya K. and Dutta, Achintya Kumar},
    title = {A reduced cost four-component relativistic unitary coupled cluster method for atoms and molecules},
    journal = { J. Chem. Phys.},
    volume = {161},
    number = {3},
    pages = {034101},
    year = {2024},
    month = {07},
    issn = {0021-9606},
    doi = {10.1063/5.0207091},
    url = {https://doi.org/10.1063/5.0207091},

}

@article{eliav1994open,
  title={Open-shell relativistic coupled-cluster method with Dirac-Fock-Breit wave functions: Energies of the gold atom and its cation},
  author={Eliav, Ephraim and Kaldor, Uzi and Ishikawa, Yasuyuki},
  journal={Physical Review A},
  volume={49},
  number={3},
  pages={1724},
  year={1994},
  publisher={APS},
  url={https://journals.aps.org/pra/abstract/10.1103/PhysRevA.49.1724}
}

@article{visscher1995kramers,
  title={Kramers-restricted closed-shell CCSD theory},
  author={Visscher, Lucas and Dyall, Kenneth G and Lee, Timothy J},
  journal={Int. J. Quantum Chem.},
  volume={56},
  number={S29},
  pages={411--419},
  year={1995},
  publisher={Wiley Online Library},
  url={https://onlinelibrary.wiley.com/doi/abs/10.1002/qua.560560844}
}

@article{eliav1994relativistic,
  title={Relativistic coupled cluster theory based on the no-pair dirac--coulomb--breit hamiltonian: Relativistic pair correlation energies of the xe atom},
  author={Eliav, Ephraim and Kaldor, Uzi and Ishikawa, Yasuyuki},
  journal={Int. J. Quantum Chem.},
  volume={52},
  number={S28},
  pages={205--214},
  year={1994},
  publisher={Wiley Online Library},
  url={https://onlinelibrary.wiley.com/doi/abs/10.1002/qua.560520821}
}

@article{visscher1996formulation,
  title={Formulation and implementation of a relativistic unrestricted coupled-cluster method including noniterative connected triples},
  author={Visscher, Lucas and Lee, Timothy J and Dyall, Kenneth G},
  journal={J. Chem. Phys. },
  volume={105},
  number={19},
  pages={8769--8776},
  year={1996},
  publisher={AIP Publishing},
  url={https://pubs.aip.org/aip/jcp/article/105/19/8769/479580/Formulation-and-implementation-of-a-relativistic}
}

@article{lee1998spin,
  title={Spin-orbit effects calculated by two-component coupled-cluster methods: test calculations on AuH, Au2, TlH and Tl2},
  author={Lee, Hyo-Sug and Han, Young-Kyu and Kim, Myeong Cheol and Bae, Cheolbeom and Lee, Yoon Sup},
  journal={Chemical physics letters},
  volume={293},
  number={1-2},
  pages={97--102},
  year={1998},
  publisher={Elsevier},
  url={https://www.sciencedirect.com/science/article/pii/S000926149800760X}
}

@article{nataraj2010general,
  title={General implementation of the relativistic coupled-cluster method},
  author={Nataraj, Huliyar S and K{\'a}llay, Mih{\'a}ly and Visscher, Lucas},
  journal={J. Chem. Phys. },
  volume={133},
  number={23},
  year={2010},
  publisher={AIP Publishing},
  url={https://pubs.aip.org/aip/jcp/article/133/23/234109/950148/General-implementation-of-the-relativistic-coupled}
}

@article{visscher2001formulation,
  title={Formulation and implementation of the relativistic Fock-space coupled cluster method for molecules},
  author={Visscher, Lucas and Eliav, Ephraim and Kaldor, Uzi},
  journal={J. Chem. Phys. },
  volume={115},
  number={21},
  pages={9720--9726},
  year={2001},
  publisher={American Institute of Physics},
  url={https://pubs.aip.org/aip/jcp/article/115/21/9720/453336/Formulation-and-implementation-of-the-relativistic}
}

@article{koulias2019relativistic,
  title={Relativistic real-time time-dependent equation-of-motion coupled-cluster},
  author={Koulias, Lauren N and Williams-Young, David B and Nascimento, Daniel R and DePrince III, A Eugene and Li, Xiaosong},
  journal={Journal of chemical theory and computation},
  volume={15},
  number={12},
  pages={6617--6624},
  year={2019},
  publisher={ACS Publications},
  url={https://pubs.acs.org/doi/10.1021/acs.jctc.9b00729}
}

@article{liu2021relativistic,
  title={Relativistic coupled-cluster and equation-of-motion coupled-cluster methods},
  author={Liu, Junzi and Cheng, Lan},
  journal={Wiley Interdisciplinary Reviews: Computational Molecular Science},
  volume={11},
  number={6},
  pages={e1536},
  year={2021},
  publisher={Wiley Online Library},
  url={https://wires.onlinelibrary.wiley.com/doi/abs/10.1002/wcms.1536}
}

@article{shee2016analytic,
  title={Analytic one-electron properties at the 4-component relativistic coupled cluster level with inclusion of spin-orbit coupling},
  author={Shee, Avijit and Visscher, Lucas and Saue, Trond},
  journal={J. Chem. Phys. },
  volume={145},
  number={18},
  year={2016},
  publisher={AIP Publishing},
  url={https://doi.org/10.1063/1.4966643}
}

@article{liu2021analytic,
  title={Analytic evaluation of energy first derivatives for spin--orbit coupled-cluster singles and doubles augmented with noniterative triples method: General formulation and an implementation for first-order properties},
  author={Liu, Junzi and Zheng, Xuechen and Asthana, Ayush and Zhang, Chaoqun and Cheng, Lan},
  journal={J. Chem. Phys. },
  volume={154},
  number={6},
  year={2021},
  publisher={AIP Publishing},
  url = {https://doi.org/10.1063/5.0038779}
}

@article{zheng2022geometry,
  title={Geometry optimizations with spinor-based relativistic coupled-cluster theory},
  author={Zheng, Xuechen and Zhang, Chaoqun and Liu, Junzi and Cheng, Lan},
  journal={J. Chem. Phys. },
  volume={156},
  number={15},
  year={2022},
  publisher={AIP Publishing},
  url = {https://doi.org/10.1063/5.0086281}
}

@article{lowdin1955quantum,
  title={Quantum theory of many-particle systems. I. Physical interpretations by means of density matrices, natural spin-orbitals, and convergence problems in the method of configurational interaction},
  author={L{\"o}wdin, Per-Olov},
  journal={Physical Review},
  volume={97},
  number={6},
  pages={1474},
  year={1955},
  publisher={APS},
  url = {https://link.aps.org/doi/10.1103/PhysRev.97.1474}
}

@article{yuan2024formulation,
  title={Formulation and Implementation of Frequency-Dependent Linear Response Properties with Relativistic Coupled Cluster Theory for GPU-accelerated Computer Architectures},
  author={Yuan, Xiang and Halbert, Lo{\"\i}c and Pototschnig, Johann Valentin and Papadopoulos, Anastasios and Coriani, Sonia and Visscher, Lucas and Pereira Gomes, André Severo},
  journal={Journal of chemical theory and computation},
  volume={20},
  number={2},
  pages={677--694},
  year={2024},
  publisher={ACS Publications},
  url={https://doi.org/10.1021/acs.jctc.3c00812}
}

@article{yuan2023frequency,
  title={Frequency-Dependent Quadratic Response Properties and Two-photon Absorption from Relativistic Equation-of-Motion Coupled Cluster Theory},
  author={Yuan, Xiang and Halbert, Lo{\"\i}c and Visscher, Lucas and Pereira Gomes, André Severo},
  journal={Journal of Chemical Theory and Computation},
  volume={19},
  number={24},
  pages={9248--9259},
  year={2023},
  publisher={ACS Publications},
  url={https://doi.org/10.1021/acs.jctc.3c01011}
}

@article{chakraborty2024spin,
author = {Chakraborty, Sudipta and Mukhopadhyay, Tamoghna and Dutta, Achintya Kumar},
title = {Spin-Free Exact Two-Component Linear Response Coupled Cluster Theory for the Estimation of Frequency-Dependent Second-Order Properties},
journal = {J. Phys. Chem. A},
doi = {10.1021/acs.jpca.4c03584},
    note ={PMID: 40152233},
URL = { https://doi.org/10.1021/acs.jpca.4c03584}
}

@article{crawford2019reduced,
  title={Reduced-scaling coupled cluster response theory: Challenges and opportunities},
  author={Crawford, T Daniel and Kumar, Ashutosh and Bazant{\'e}, Alexandre P and Di Remigio, Roberto},
  journal={Wiley Interdisciplinary Reviews: Computational Molecular Science},
  volume={9},
  number={4},
  pages={e1406},
  year={2019},
  publisher={Wiley Online Library},
  url = {https://wires.onlinelibrary.wiley.com/doi/abs/10.1002/wcms.1406}
}

@article{olsen1985linear,
  title={Linear and nonlinear response functions for an exact state and for an MCSCF state},
  author={Olsen, Jeppe and Jo/rgensen, Poul},
  journal={J. Chem. Phys. },
  volume={82},
  number={7},
  pages={3235--3264},
  year={1985},
  publisher={American Institute of Physics},
  url = {https://doi.org/10.1063/1.448223},
}

@article{goebel1995dispersion,
  title={Dispersion of the refractive index of cadmium vapor and the dipole polarizability of the atomic cadmium 1 S 0 state},
  author={Goebel, Dirk and Hohm, Uwe},
  journal={Physical Review A},
  volume={52},
  number={5},
  pages={3691},
  year={1995},
  publisher={APS},
  URL={https://doi.org/10.1103/PhysRevA.52.3691}
}

@article{goebel1996theoretical,
  title = {Theoretical and experimental determination of the polarizabilities of the zinc $^{1}$${\mathit{S}}_{0}$ state},
  author = {Goebel, Dirk and Hohm, Uwe and Maroulis, George},
  journal = {Phys. Rev. A},
  volume = {54},
  issue = {3},
  pages = {1973--1978},
  numpages = {0},
  year = {1996},
  month = {Sep},
  publisher = {American Physical Society},
  doi = {10.1103/PhysRevA.54.1973},
  url = {https://link.aps.org/doi/10.1103/PhysRevA.54.1973}
}

@article{goebel1996dipole,
  title={Dipole polarizability, Cauchy moments, and related properties of Hg},
  author={Goebel, Dirk and Hohm, Uwe},
  journal={The Journal of Physical Chemistry},
  volume={100},
  number={18},
  pages={7710--7712},
  year={1996},
  publisher={ACS Publications},
  URL={https://doi.org/10.1021/jp960231l}
}

@article{hohm2022dipole,
  title={Dipole--Dipole Polarizability of the Cadmium 1 S 0 State Revisited},
  author={Hohm, Uwe},
  journal={Optics and Spectroscopy},
  volume={130},
  number={4},
  pages={290--294},
  year={2022},
  publisher={Springer},
  URL={https://doi.org/10.1134/S0030400X22040105}
}

@article{dutta2023bagh,
  title = {{{BAGH}}: {{A}} Quantum Chemistry Software Package},
  author = {Dutta, Achintya Kumar and Manna, Amrita and Jangid, Bhavnesh and Majee, Kamal and Surjuse, Kshitijkumar and Mukherjee, Madhubani and Thapa, Mrinal and Begom, Muskan and Arora, Sneha and Chamoli, Somesh and Haldar, Soumi and Chakraborty, Sudipta and Mandal, Sujan and Mukhopadhyay, Tamoghna and Bhattacharjya, Violina},
  year = {2025},
  month = feb,
  URL = {https://bagh-doc.readthedocs.io/en/latest/},
  urldate = {2025-02-21},
}

@article{hohm2013experimental,
  title={Experimental static dipole--dipole polarizabilities of molecules},
  author={Hohm, Uwe},
  journal={Journal of Molecular Structure},
  volume={1054},
  pages={282--292},
  year={2013},
  publisher={Elsevier},
  URL={https://doi.org/10.1016/j.molstruc.2013.10.003}
}

@article{maroulis1997electrooptical,
  title={Electrooptical properties and molecular polarization of iodine, I2},
  author={Maroulis, George and Makris, Constantinos and Hohm, Uwe and Goebel, Dirk},
  journal={J. Phys. Chem. A},
  volume={101},
  number={5},
  pages={953--956},
  year={1997},
  publisher={ACS Publications},
  URL={https://doi.org/10.1021/jp962578u}
  
}

@article{pyscf2020,
    author = {Sun, Qiming and Zhang, Xing and Banerjee, Samragni and Bao, Peng and Barbry, Marc and Blunt, Nick S. and Bogdanov, Nikolay A. and Booth, George H. and Chen, Jia and Cui, Zhi-Hao and Eriksen, Janus J. and Gao, Yang and Guo, Sheng and Hermann, Jan and Hermes, Matthew R. and Koh, Kevin and Koval, Peter and Lehtola, Susi and Li, Zhendong and Liu, Junzi and Mardirossian, Narbe and McClain, James D. and Motta, Mario and Mussard, Bastien and Pham, Hung Q. and Pulkin, Artem and Purwanto, Wirawan and Robinson, Paul J. and Ronca, Enrico and Sayfutyarova, Elvira R. and Scheurer, Maximilian and Schurkus, Henry F. and Smith, James E. T. and Sun, Chong and Sun, Shi-Ning and Upadhyay, Shiv and Wagner, Lucas K. and Wang, Xiao and White, Alec and Whitfield, James Daniel and Williamson, Mark J. and Wouters, Sebastian and Yang, Jun and Yu, Jason M. and Zhu, Tianyu and Berkelbach, Timothy C. and Sharma, Sandeep and Sokolov, Alexander Yu. and Chan, Garnet Kin-Lic},
    title = {Recent developments in the PySCF program package},
    journal = {The Journal of Chemical Physics},
    volume = {153},
    number = {2},
    pages = {024109},
    year = {2020},
    month = {07},
    issn = {0021-9606},
    doi = {10.1063/5.0006074},
    url = {https://doi.org/10.1063/5.0006074},
}

@article{Qiming2015,
author = {Sun, Qiming},
title = {Libcint: An efficient general integral library for Gaussian basis functions},
journal = {Journal of Computational Chemistry},
volume = {36},
number = {22},
pages = {1664-1671},
doi = {https://doi.org/10.1002/jcc.23981},
url = {https://onlinelibrary.wiley.com/doi/abs/10.1002/jcc.23981},
year = {2015}
}

@article{Qiming2018,
author = {Sun, Qiming and Berkelbach, Timothy C. and Blunt, Nick S. and Booth, George H. and Guo, Sheng and Li, Zhendong and Liu, Junzi and McClain, James D. and Sayfutyarova, Elvira R. and Sharma, Sandeep and Wouters, Sebastian and Chan, Garnet Kin-Lic},
title = {PySCF: the Python-based simulations of chemistry framework},
journal = {WIREs Computational Molecular Science},
volume = {8},
number = {1},
pages = {e1340},
doi = {https://doi.org/10.1002/wcms.1340},
url = {https://wires.onlinelibrary.wiley.com/doi/abs/10.1002/wcms.1340},
year = {2018}
}

@article{Barca2020,
    author = {Barca, Giuseppe M. J. and Bertoni, Colleen and Carrington, Laura and Datta, Dipayan and De Silva, Nuwan and Deustua, J. Emiliano and Fedorov, Dmitri G. and Gour, Jeffrey R. and Gunina, Anastasia O. and Guidez, Emilie and Harville, Taylor and Irle, Stephan and Ivanic, Joe and Kowalski, Karol and Leang, Sarom S. and Li, Hui and Li, Wei and Lutz, Jesse J. and Magoulas, Ilias and Mato, Joani and Mironov, Vladimir and Nakata, Hiroya and Pham, Buu Q. and Piecuch, Piotr and Poole, David and Pruitt, Spencer R. and Rendell, Alistair P. and Roskop, Luke B. and Ruedenberg, Klaus and Sattasathuchana, Tosaporn and Schmidt, Michael W. and Shen, Jun and Slipchenko, Lyudmila and Sosonkina, Masha and Sundriyal, Vaibhav and Tiwari, Ananta and Galvez Vallejo, Jorge L. and Westheimer, Bryce and Włoch, Marta and Xu, Peng and Zahariev, Federico and Gordon, Mark S.},
    title = {Recent developments in the general atomic and molecular electronic structure system},
    journal = {The Journal of Chemical Physics},
    volume = {152},
    number = {15},
    pages = {154102},
    year = {2020},
    month = {04},
    issn = {0021-9606},
    doi = {10.1063/5.0005188},
    url = {https://doi.org/10.1063/5.0005188},
}

@article{DIRAC_saue2020,
    author = {Saue, Trond and Bast, Radovan and Gomes, André Severo Pereira and Jensen, Hans Jørgen Aa. and Visscher, Lucas and Aucar, Ignacio Agustín and Di Remigio, Roberto and Dyall, Kenneth G. and Eliav, Ephraim and Fasshauer, Elke and Fleig, Timo and Halbert, Loïc and Hedegård, Erik Donovan and Helmich-Paris, Benjamin and Iliaš, Miroslav and Jacob, Christoph R. and Knecht, Stefan and Laerdahl, Jon K. and Vidal, Marta L. and Nayak, Malaya K. and Olejniczak, Małgorzata and Olsen, Jógvan Magnus Haugaard and Pernpointner, Markus and Senjean, Bruno and Shee, Avijit and Sunaga, Ayaki and van Stralen, Joost N. P.},
    title = {The DIRAC code for relativistic molecular calculations},
    journal = {The Journal of Chemical Physics},
    volume = {152},
    number = {20},
    pages = {204104},
    year = {2020},
    month = {05},
    issn = {0021-9606},
    doi = {10.1063/5.0004844},
    url = {https://doi.org/10.1063/5.0004844},
    eprint = {https://pubs.aip.org/aip/jcp/article-pdf/doi/10.1063/5.0004844/16745096/204104\_1\_online.pdf},
}

@article{hess1986relativistic,
  title={Relativistic electronic-structure calculations employing a two-component no-pair formalism with external-field projection operators},
  author={Hess, Bernd A},
  journal={Physical Review A},
  volume={33},
  number={6},
  pages={3742},
  year={1986},
  publisher={APS},
  url = {https://link.aps.org/doi/10.1103/PhysRevA.33.3742}
}

@article{van1996relativistic,
  title={Relativistic regular two-component Hamiltonians},
  author={van Lenthe, Erik and Van Leeuwen, R and Baerends, EJ and Snijders, JG},
  journal={International Journal of Quantum Chemistry},
  volume={57},
  number={3},
  pages={281--293},
  year={1996},
  publisher={Wiley Online Library},
  url = {https://onlinelibrary.wiley.com/doi/abs/10.1002/%28SICI%291097-461X%281996%2957%3A3%3C281%3A%3AAID-QUA2%3E3.0.CO%3B2-U}
}

@article{dyall1997interfacing,
  title={Interfacing relativistic and nonrelativistic methods. I. Normalized elimination of the small component in the modified Dirac equation},
  author={Dyall, Kenneth G},
  journal={The Journal of chemical physics},
  volume={106},
  number={23},
  pages={9618--9626},
  year={1997},
  publisher={American Institute of Physics},
  url = {https://doi.org/10.1063/1.473860}
}

@article{nakajima1999new,
  title={A new relativistic theory: a relativistic scheme by eliminating small components (RESC)},
  author={Nakajima, Takahito and Hirao, Kimihiko},
  journal={Chemical physics letters},
  volume={302},
  number={5-6},
  pages={383--391},
  year={1999},
  publisher={Elsevier},
  url = {https://www.sciencedirect.com/science/article/pii/S0009261499001505}
}

@article{barysz2001two,
  title={Two-component methods of relativistic quantum chemistry: from the Douglas--Kroll approximation to the exact two-component formalism},
  author={Barysz, Maria and Sadlej, Andrzej J},
  journal={Journal of Molecular Structure: THEOCHEM},
  volume={573},
  number={1-3},
  pages={181--200},
  year={2001},
  publisher={Elsevier},
  url = {https://www.sciencedirect.com/science/article/pii/S0166128001005425}
}

@article{liu2009exact,
  title={Exact two-component Hamiltonians revisited},
  author={Liu, Wenjian and Peng, Daoling},
  journal={The Journal of chemical physics},
  volume={131},
  number={3},
  year={2009},
  publisher={AIP Publishing},
  url = {https://doi.org/10.1063/1.3159445}
}

@article{saue2011relativistic1,
  title={Relativistic Hamiltonians for chemistry: A primer},
  author={Saue, Trond},
  journal={ChemPhysChem},
  volume={12},
  number={17},
  pages={3077--3094},
  year={2011},
  url = {https://chemistry-europe.onlinelibrary.wiley.com/doi/abs/10.1002/cphc.201100682}
}

@article{helmich2019relativistic,
  title={Relativistic Cholesky-decomposed density matrix MP2},
  author={Helmich-Paris, Benjamin and Repisky, Michal and Visscher, Lucas},
  journal={Chemical Physics},
  volume={518},
  pages={38--46},
  year={2019},
  publisher={Elsevier},
  url = {https://www.sciencedirect.com/science/article/pii/S0301010418311388}
}

@article{banerjee2023relativistic,
  title={Relativistic resolution-of-the-identity with Cholesky integral decomposition},
  author={Banerjee, Samragni and Zhang, Tianyuan and Dyall, Kenneth G and Li, Xiaosong},
  journal={The Journal of Chemical Physics},
  volume={159},
  number={11},
  year={2023},
  publisher={AIP Publishing},
  url = {https://doi.org/10.1063/5.0161871}
}

@article{uhlirova2024cholesky,
  title={Cholesky Decomposition in Spin-Free Dirac--Coulomb Coupled-Cluster Calculations},
  author={Uhlirova, Tereza and Cianchino, Davide and Nottoli, Tommaso and Lipparini, Filippo and Gauss, Jurgen},
  journal={The Journal of Physical Chemistry A},
  volume={128},
  number={38},
  pages={8292--8303},
  year={2024},
  publisher={ACS Publications},
  url = {https://doi.org/10.1021/acs.jpca.4c04353}
}

@article{van2005accurate,
  title={Accurate and efficient treatment of two-electron contributions in quasirelativistic high-order Douglas-Kroll density-functional calculations},
  author={Van W{\"u}llen, Christoph and Michauk, Christine},
  journal={The Journal of chemical physics},
  volume={123},
  number={20},
  year={2005},
  publisher={AIP Publishing}, 
  url = {https://doi.org/10.1063/1.2133731}
}

@article{liu2006infinite,
  title={Infinite-order quasirelativistic density functional method based on the exact matrix quasirelativistic theory},
  author={Liu, Wenjian and Peng, Daoling},
  journal={The Journal of chemical physics},
  volume={125},
  number={4},
  year={2006},
  publisher={AIP Publishing}, 
  url = {https://doi.org/10.1063/1.2222365}
}

@article{peng2007making,
  title={Making four-and two-component relativistic density functional methods fully equivalent based on the idea of “from atoms to molecule”},
  author={Peng, Daoling and Liu, Wenjian and Xiao, Yunlong and Cheng, Lan},
  journal={The Journal of chemical physics},
  volume={127},
  number={10},
  year={2007},
  publisher={AIP Publishing},
  url = {https://doi.org/10.1063/1.2772856}
}

@article{knecht,
  title={Exact two-component Hamiltonians for relativistic quantum chemistry: Two-electron picture-change corrections made simple},
  author={Knecht, Stefan and Repisky, Michal and Jensen, Hans J{\o}rgen Aagaard and Saue, Trond},
  journal={The Journal of Chemical Physics},
  volume={157},
  number={11},
  year={2022},
  publisher={AIP Publishing},
  url = {https://doi.org/10.1063/5.0095112}
}

@article{samzow1992two,
  title={The two-electron terms of the no-pair Hamiltonian},
  author={Samzow, Reinhard and Hess, Bernd A and Jansen, Georg},
  journal={The Journal of chemical physics},
  volume={96},
  number={2},
  pages={1227--1231},
  year={1992},
  publisher={American Institute of Physics}
}

@article{ikabata2021picture,
  title={Picture-change correction in relativistic density functional theory},
  author={Ikabata, Yasuhiro and Nakai, Hiromi},
  journal={Physical Chemistry Chemical Physics},
  volume={23},
  number={29},
  pages={15458--15474},
  year={2021},
  publisher={Royal Society of Chemistry}
}

@article{chakraborty2025spin,
  title={Spin-Free Exact Two-Component Linear Response Coupled Cluster Theory for the Estimation of Frequency-Dependent Second-Order Properties},
  author={Chakraborty, Sudipta and Mukhopadhyay, Tamoghna and Dutta, Achintya Kumar},
  journal={The Journal of Physical Chemistry A},
  volume={129},
  number={14},
  pages={3315--3330},
  year={2025},
  publisher={ACS Publications},
  url = {https://doi.org/10.1021/acs.jpca.4c03584}
}

@article{chakraborty2025low-cost,
    author = {Chakraborty, Sudipta and Manna, Amrita and Crawford, T. Daniel and Dutta, Achintya Kumar},
    title = {A low-cost four-component relativistic coupled cluster linear response theory based on perturbation sensitive natural spinors},
    journal = {The Journal of Chemical Physics},
    volume = {163},
    number = {4},
    pages = {044111},
    year = {2025},
    month = {07},
    issn = {0021-9606},
    doi = {10.1063/5.0274450},
    url = {https://doi.org/10.1063/5.0274450},
}

@article{dalgaard1980time,
  title={Time-dependent multiconfigurational Hartree--Fock theory},
  author={Dalgaard, Esper},
  journal={The Journal of Chemical Physics},
  volume={72},
  number={2},
  pages={816--823},
  year={1980},
  publisher={AIP Publishing},
  url = {https://doi.org/10.1063/1.439233}
}

@article{rice1991calculation,
  title={The calculation of frequency-dependent polarizabilities as pseudo-energy derivatives},
  author={Rice, Julia E and Handy, Nicholas C},
  journal={The Journal of chemical physics},
  volume={94},
  number={7},
  pages={4959--4971},
  year={1991},
  publisher={American Institute of Physics},
  url = {https://doi.org/10.1063/1.460558}
}

@article{saue1997principles,
  title={Principles of direct 4-component relativistic SCF: application to caesium auride},
  author={Saue, By T and F{\ae}gri, Knut and Helgaker, Trygve and Gropen, Odd},
  journal={Molecular Physics},
  volume={91},
  number={5},
  pages={937--950},
  year={1997},
  publisher={Taylor \& Francis},
  url = {https://www.tandfonline.com/doi/abs/10.1080/002689797171058}
}

@article{klopper1997multiple,
  title={Multiple basis sets in calculations of triples corrections in coupled-cluster theory},
  author={Klopper, Wim and Noga, Jozef and Koch, Henrik and Helgaker, Trygve},
  journal={Theoretical Chemistry Accounts},
  volume={97},
  number={1},
  pages={164--176},
  year={1997},
  publisher={Springer},
  url = {https://doi.org/10.1007/s002140050250}
}

@article{nalewajski1995proceedings,
  title={Proceedings of the satellite symposium on “thirty years of density functional theory: Concepts and applications”(Cracow, June 13--16, 1994)},
  author={Nalewajski, Roman F},
  journal={International Journal of Quantum Chemistry},
  volume={56},
  number={4},
  pages={197--198},
  year={1995},
  publisher={Wiley Online Library},
  url = {https://onlinelibrary.wiley.com/doi/abs/10.1002/qua.560560402}
}

@article{peric1996theoretical,
  title={On a theoretical model for the Renner--Teller effect in tetra-atomic molecules},
  author={Peri{\'c}, Miljenko and Ostoji{\'c}, B and Engels, B},
  journal={The Journal of chemical physics},
  volume={105},
  number={19},
  pages={8569--8585},
  year={1996},
  publisher={American Institute of Physics},
  url = {https://doi.org/10.1063/1.472641}
}

@article{el2005theoretical,
  title={Theoretical study of H3AXH3 and H3AYH2 (A= B, Al, Ga; X= N, P, As and Y= O, S, and Se), electrostatic and hyperconjugative interactions roles},
  author={El Guerraze, Abdela{\^a}li and El-Nahas, Ahmed M and Jarid, Abdellah and Serrar, Chafiq and Anane, Hafid and Esseffar, M’hamed},
  journal={Chemical physics},
  volume={313},
  number={1-3},
  pages={159--168},
  year={2005},
  publisher={Elsevier},
  url = {https://www.sciencedirect.com/science/article/pii/S0301010405000108}
}

@article{mandal2026third,
  title={A third-order relativistic algebraic diagrammatic construction method for double ionization potentials: Theory, implementation, and benchmark},
  author={Mandal, Sujan and Dutta, Achintya Kumar},
  journal={The Journal of Chemical Physics},
  volume={164},
  number={4},
  year={2026},
  publisher={AIP Publishing},
  url = {https://doi.org/10.1063/5.0305197}
}

@article{mukhopadhyay2025reduced,
  title={Reduced-cost relativistic equation-of-motion coupled cluster method based on frozen natural spinors: A state-specific approach},
  author={Mukhopadhyay, Tamoghna and Thapa, Mrinal and Chamoli, Somesh and Wang, Xubo and Zhang, Chaoqun and Nayak, Malaya K and Dutta, Achintya Kumar},
  journal={The Journal of Chemical Physics},
  volume={163},
  number={19},
  year={2025},
  publisher={AIP Publishing},
  url = {https://doi.org/10.1063/5.0289155}
}

@article{chakraborty2025low,
  title={A Low Cost Relativistic Algebraic Diagrammatic Construction Method Based on Cholesky Decomposition and Frozen Natural Spinors for Electronic Ionization, Attachment and Excitation Energy Problem},
  author={Chakraborty, Sudipta and Majee, Kamal and Dutta, Achintya Kumar},
  journal={Journal of Chemical Theory and Computation},
  year={2025},
  publisher={ACS Publications},
  url = {https://doi.org/10.1021/acs.jctc.6c00127}
}

@article{wang2025relativistic,
  title={Relativistic two-electron contributions within exact two-component theory},
  author={Wang, Xubo and Zhang, Chaoqun and Liu, Junzi and Cheng, Lan},
  journal={Chemical Physics Reviews},
  volume={6},
  number={3},
  year={2025},
  publisher={AIP Publishing},
  url = {https://doi.org/10.1063/5.0268348}
}

@article{knechtExactTwocomponentHamiltonians2022,
  title = {Exact Two-Component {{Hamiltonians}} for Relativistic Quantum Chemistry: {{Two-electron}} Picture-Change Corrections Made Simple},
  shorttitle = {Exact Two-Component {{Hamiltonians}} for Relativistic Quantum Chemistry},
  author = {Knecht, Stefan and Repisky, Michal and Jensen, Hans J{\o}rgen Aagaard and Saue, Trond},
  year = {2022},
  month = sep,
  journal = {The Journal of Chemical Physics},
  volume = {157},
  number = {11},
  pages = {114106},
  issn = {0021-9606},
  doi = {10.1063/5.0095112},
  urldate = {2025-01-01}
}

@article{zhangAtomicMeanFieldApproach2022,
  title = {Atomic {{Mean-Field Approach}} within {{Exact Two-Component Theory Based}} on the {{Dirac}}--{{Coulomb}}--{{Breit Hamiltonian}}},
  author = {Zhang, Chaoqun and Cheng, Lan},
  year = {2022},
  month = jul,
  journal = {The Journal of Physical Chemistry A},
  volume = {126},
  number = {27},
  pages = {4537--4553},
  publisher = {American Chemical Society},
  issn = {1089-5639},
  doi = {10.1021/acs.jpca.2c02181},
  urldate = {2025-01-01}
}

@incollection{aquilante2011cholesky,
  title={Cholesky decomposition techniques in electronic structure theory},
  author={Aquilante, Francesco and Boman, Linus and Bostr{\"o}m, Jonas and Koch, Henrik and Lindh, Roland and de Mer{\'a}s, Alfredo S{\'a}nchez and Pedersen, Thomas Bondo},
  booktitle={Linear-Scaling Techniques in Computational Chemistry and Physics: Methods and Applications},
  pages={301--343},
  year={2011},
  publisher={Springer},
  url = {https://doi.org/10.1007/978-90-481-2853-2_13}
}

@article{folkestad2019efficient,
  title={An efficient algorithm for Cholesky decomposition of electron repulsion integrals},
  author={Folkestad, Sarai D and Kj{\o}nstad, Eirik F and Koch, Henrik},
  journal={The Journal of chemical physics},
  volume={150},
  number={19},
  year={2019},
  publisher={AIP Publishing},
  url = {https://doi.org/10.1063/1.5083802}
}

@article{zhang2021toward,
  title={Toward the minimal floating operation count Cholesky decomposition of electron repulsion integrals},
  author={Zhang, Tianyuan and Liu, Xiaolin and Valeev, Edward F and Li, Xiaosong},
  journal={The Journal of Physical Chemistry A},
  volume={125},
  number={19},
  pages={4258--4265},
  year={2021},
  publisher={ACS Publications},
  url = {https://doi.org/10.1021/acs.jpca.1c02317}
}

@article{liu2018,
  title={An atomic mean-field spin-orbit approach within exact two-component theory for a non-perturbative treatment of spin-orbit coupling},
  author={Liu, Junzi and Cheng, Lan},
  journal={The Journal of Chemical Physics},
  volume={148},
  number={14},
  year={2018},
  publisher={AIP Publishing},
  url = {https://doi.org/10.1063/1.5023750}
}

@article{Zhang2022,
author = {Zhang, Chaoqun and Cheng, Lan},
title = {Atomic Mean-Field Approach within Exact Two-Component Theory Based on the Dirac-Coulomb-Breit Hamiltonian},
journal = {The Journal of Physical Chemistry A},
volume = {126},
number = {27},
pages = {4537-4553},
year = {2022},
doi = {10.1021/acs.jpca.2c02181},
URL = {https://doi.org/10.1021/acs.jpca.2c02181
},

}

@article{wolf2002generalized,
  title={The generalized douglas--kroll transformation},
  author={Wolf, Alexander and Reiher, Markus and Hess, Bernd Artur},
  journal={The Journal of chemical physics},
  volume={117},
  number={20},
  pages={9215--9226},
  year={2002},
  publisher={American Institute of Physics}
}

@article{nakajima2000higher,
  title={The higher-order Douglas--Kroll transformation},
  author={Nakajima, Takahito and Hirao, Kimihiko},
  journal={The Journal of Chemical Physics},
  volume={113},
  number={18},
  pages={7786--7789},
  year={2000},
  publisher={American Institute of Physics}
}

@article{reiher2004exactI,
  title={Exact decoupling of the Dirac Hamiltonian. I. General theory},
  author={Reiher, Markus and Wolf, Alexander},
  journal={The Journal of chemical physics},
  volume={121},
  number={5},
  pages={2037--2047},
  year={2004},
  publisher={American Institute of Physics}
}

@article{reiher2004exactII,
  title={Exact decoupling of the Dirac Hamiltonian. II. The generalized Douglas--Kroll--Hess transformation up to arbitrary order},
  author={Reiher, Markus and Wolf, Alexander},
  journal={The Journal of chemical physics},
  volume={121},
  number={22},
  pages={10945--10956},
  year={2004},
  publisher={AIP Publishing}
}

@article{peng2009arbitrary,
  title={An arbitrary order Douglas--Kroll method with polynomial cost},
  author={Peng, Daoling and Hirao, Kimihiko},
  journal={The Journal of chemical physics},
  volume={130},
  number={4},
  year={2009},
  publisher={AIP Publishing}
}

@article{barysz1997expectation,
  title={Expectation values of operators in approximate two-component relativistic theories},
  author={Barysz, Maria and Sadlej, Andrzej J},
  journal={Theoretical Chemistry Accounts},
  volume={97},
  number={1},
  pages={260--270},
  year={1997},
  publisher={Springer}
}

@article{barysz2002infinite,
  title={Infinite-order two-component theory for relativistic quantum chemistry},
  author={Barysz, Maria and Sadlej, Andrzej J},
  journal={The Journal of chemical physics},
  volume={116},
  number={7},
  pages={2696--2704},
  year={2002},
  publisher={American Institute of Physics}
}

@article{dyall2001interfacing,
  title={Interfacing relativistic and nonrelativistic methods. IV. One-and two-electron scalar approximations},
  author={Dyall, Kenneth G},
  journal={The Journal of Chemical Physics},
  volume={115},
  number={20},
  pages={9136--9143},
  year={2001},
  publisher={American Institute of Physics}
}

@article{cheng2011analytic,
  title={Analytic energy gradients for the spin-free exact two-component theory using an exact block diagonalization for the one-electron Dirac Hamiltonian},
  author={Cheng, Lan and Gauss, J{\"u}rgen},
  journal={The Journal of chemical physics},
  volume={135},
  number={8},
  year={2011},
  publisher={AIP Publishing}
}

@article{sikkema2009molecular,
  title={The molecular mean-field approach for correlated relativistic calculations},
  author={Sikkema, Jetze and Visscher, Lucas and Saue, Trond and Ilia{\v{s}}, Miroslav},
  journal={The Journal of chemical physics},
  volume={131},
  number={12},
  year={2009},
  publisher={AIP Publishing}
}

\end{document}



\begin{center}

{\bfseries Supplementary Information}

\vspace{1cm}

{\bfseries
Efficient Implementation of Relativistic Coupled Cluster Linear Response Theory in Combination with Perturbation Sensitive Natural Spinors and Cholesky Decomposition Treatment of Two-electron Integrals
\par}

\vspace{0.8cm}

{
Sudipta Chakraborty$^{1}$,
Muskan Begom$^{1}$,
Xubo Wang$^{2}$,
Achintya Kumar Dutta$^{1,*}$
\par}

\vspace{0.6cm}

{\itshape
$^{1}$Department of Chemistry, Indian Institute of Technology Bombay,
Mumbai 400076, India
\par}

\vspace{0.25cm}

{\itshape
$^{2}$Department of Chemistry,
The Johns Hopkins University,
Baltimore, Maryland 21218, USA
\par}

\vspace{0.5cm}

{\small
$^{*}$Corresponding author:
\href{mailto:achintya@chem.iitb.ac.in}{achintya@chem.iitb.ac.in}
}

\end{center}

\vspace{1cm}


\clearpage


\begin{figure}[h]
\centering
    \begin{subfigure}{0.8\textwidth} 
        \includegraphics[width=\linewidth]{figures/cd_three_pannel_4c_amf_mp.jpg} 
    \end{subfigure}

    \caption{\label{fig:cd_dispersion} Polarizability spectra of  Cd, using relativistic LRCC using 4c-DC, X2CAMF and X2CMP Hamiltonian with the s-aug-dyall.v2z basis set.}
\end{figure}

\begin{figure}[h]
\centering
    \begin{subfigure}{0.8\textwidth} 
        \includegraphics[width=\linewidth]{figures/hg_three_pannel_4c_amf_mp.jpg} 
    \end{subfigure}

    \caption{\label{fig:hg_dispersion} Polarizability spectra of  Hg, using relativistic LRCC using 4c-DC, X2CAMF and X2CMP Hamiltonian with the s-aug-dyall.v2z basis set.}
\end{figure}

\begin{figure}[h]
\centering
    \begin{subfigure}{0.8\textwidth} 
        \includegraphics[width=\linewidth]{figures/povo_zn_dynamic_mp.jpg} 
    \end{subfigure}

    \caption{\label{fig:povo_zn_dynamic_si}Percentage of error in dynamic polarizability of the Zn atom computed with the uncontracted s-aug-dyall.v2z basis set, using FNS and FNS++ truncation schemes and X2CMP (MP) Hamiltonian, as a function of the percentage of virtual orbitals (POVO) retained.}
\end{figure}

\begin{table*}
\caption{\label{tab:diatomic}Static polarizabilities (a.u.) of some molecules using 10$^{-5}$ FNS++ occupation threshold at 4c- and X2CMP-LRCCSD/d-aug-dyall.v4z level of theory. CD threshold of 10$^{-5}$ is used. }
\begin{ruledtabular}
\begin{tabular}{rrrrrrrr} 
& \multicolumn{3}{c}{4c} &
   \multicolumn{3}{c}{X2CMP} \\
\cline{2-4} \cline{5-7} 
 &$\alpha_{\perp}$&$\alpha_{\parallel}$&$\alpha$ &$\alpha_{\perp}$&$\alpha_{\parallel}$&$\alpha$ & Expt.\cite{hohm2013experimental}\\
\hline
\\
HF         & 5.38   & 6.56   & 5.77     & 5.37   &6.55     & 5.76       & 5.60 $\pm$ 0.10 \\
HCl        & 17.07  & 18.75  & 17.56    & 17.05  &18.75    & 17.61      & 17.39 $\pm$ 0.20 \\
HBr        & 23.35  & 25.49  & 24.06    & 23.30  &25.42    & 24.00      & 23.74 $\pm$ 0.50\\
HI         & 34.78  & 37.56  & 35.71    & 34.71  &37.49    & 35.63      & 35.30 $\pm$ 0.50\\
\\  
F$_{2}$    & 6.58   & 12.52  & 8.56     & 6.55   & 12.49   & 8.53       & 8.38 $\pm$ 0.15\\ 
Cl$_{2}$   & 25.36  & 42.59  & 31.11    & 25.30  & 42.50   & 31.03      & 30.43 $\pm$ 0.30\\  
Br$_{2}$   & 36.32  & 62.84  & 45.16    & 36.24  & 62.74   & 45.07      & 47.50 $\pm$ 1.10\\ 
I$_{2}$    & 56.47  & 102.22 & 71.72    & 56.35  & 101.96  & 71.55      & 69.70 $\pm$ 1.80\\
ICl        & 41.03  & 67.01  & 49.69    & 40.95  & 66.88   & 49.59      & 43.80 $\pm$ 4.40\\
\\
AuH        & 36.94  & 41.23  & 38.37    & 36.91   &41.14  & 38.32       & \\
AuF        & 30.57  & 39.05  & 33.40    & 30.57   &39.16  & 33.43       & \\
AuCl       & 41.06  & 64.94  & 49.02    & 41.07   &64.93  & 49.02       & \\
HgCl$_{2}$ & 46.39  & 98.06  & 63.61    & 46.33   &98.01  & 63.56       & 61.20 $\pm$ 1.30\\
\end{tabular} 
\end{ruledtabular}
\end{table*}

\clearpage

\begin{table}[htbp]
\centering
\caption{Number of frozen core electrons (N$_{fc}$) used in the calculations for each atom and molecule considered in this work.}
\label{tab:frozen_electrons}

\setlength{\tabcolsep}{10pt}

\begin{tabular}{lc|lc}
\hline
System & N$_{fc}$ & System & N$_{fc}$ \\
\hline
Zn & 10  & HF    & 2  \\
Cd & 28  & HCl   & 10 \\
Hg & 46  & HBr   & 18 \\
F$_2$   & 4   & HI    & 36 \\
Cl$_2$  & 20  & AuH   & 46 \\
Br$_2$  & 36  & AuF   & 48 \\
I$_2$   & 72  & AuCl  & 56 \\
ICl     & 46  & HgCl$_2$ & 66 \\
\hline
\end{tabular}
\end{table}

\section{References}
\bibliography{references}